\documentclass[12pt,paper]{article}
\usepackage{amsmath,amssymb,amsthm,amsfonts,fullpage}
\usepackage{enumitem}  
\usepackage[
backend=bibtex,
style=numeric,
sorting=nyt
]{biblatex}
\usepackage{authblk}
\usepackage[hidelinks,pdfauthor={Abdelghaffar Chibloun}]{hyperref}
\hypersetup{
	colorlinks   = true,
	citecolor    = red
}
\usepackage{relsize}
\usepackage{cleveref}
\usepackage{color,soul}
\usepackage{color,xcolor,soul}

\soulregister\cite7
\soulregister\ref7
\providecommand{\keywords}[1]
{
	{\small	
		\textbf{{Keywords--}} #1}
}
\providecommand{\msc}[1]
{
	{\small	
		\textbf{{Mathematics Subject Classification (2020) --}} #1}
}

\theoremstyle{plain}
\newtheorem{definition}{Definition}
\newtheorem{theorem}{Theorem}
\newtheorem{proposition}{Proposition}
\newtheorem{corollary}{Corollary}
\newtheorem{lemma}{Lemma}
\newtheorem{remark}{Remark}
\newtheorem{example}{Example}
\title{{On polycyclic linear and additive codes associated to a trinomial over a finite chain ring}}
\author[1]{Abdelghaffar Chibloun\thanks{E-mail: \rm{ab.chibloun@edu.umi.ac.ma}}}
\author[2]{Hassan  Ou-azzou\thanks{E-mail: \rm{hassan.ouazzou@student.unisg.ch}. Supported by a Swiss Government Excellence Scholarship (ESKAS), no:  2024.0504.}}
\author[3]{Edgar Mart\'inez-Moro\thanks{E-mail: \rm{edgar.martinez@uva.es}. Partially supported by Grant  SAGACT-1
		MCIN/AEI/ 10.13039/501100011033 y FEDER "Una manera de hacer Europa"
		PID2022-138906NB-C21
		2023/2027}}
\author[4]{Mustapha Najmeddine\thanks{E-mail: \rm{m.najmeddine@umi.ac.ma}}}

\affil[1]{Department of Mathematics, FS–Meknes, Moulay Ismail University, Meknes, Morocco}
\affil[2]{Institute of Computer Science, University of St.Gallen, St. Gallen, Switzerland}
\affil[3]{ Institute of Mathematics, University of Valladolid, Castilla, Spain}
\affil[4]{Department of Mathematics, ENSAM–Meknes, Moulay Ismail University, Meknes, Morocco}
\date{}
\begin{document}

	\maketitle
	\begin{abstract}
		In this paper, we investigate polycyclic codes associated with a trinomial of arbitrary degree $n$ over a finite chain ring $ R.$ We extend the concepts of $ n $-isometry and $ n $-equivalence known for constacyclic codes to this class of codes, providing a broader framework for their structural analysis. We describe the classes of $n$-equivalence and compute their number, significantly reducing the study of trinomial codes over \( R\). Additionally, we examine the special case of trinomials of the form $ x^n - a_1x - a_0 \in R[x] $ and analyze their implications. Finally, we consider the extension of our results to certain trinomial additive codes over $ R.$  
		
	\end{abstract}
	
	
	\msc{94B15 - 13B25 - 81P70}
	
	\section{Introduction}

	A linear code $C$ of length $n$ over a ring $R$ is just an $R$-submodule of $R^n.$ While classical algebraic coding theory focused mainly on codes over finite fields, the discovery that many families of nonlinear codes are in fact images of linear ones over $\mathbb{Z}_4$ under a Gray map sparked an interest in codes over rings, see for example \cite{dougherty2017algebraic} for a detailed discussion of coding theory over finite commutative rings. Codes over a particular family of rings have been extensively investigated, namely those over commutative finite chain rings. One of the many reasons for this is that they are finite Frobenius rings, thus, as was shown in \cite{wood1999duality}, important theorems from classical algebraic coding theory are verified for this type of rings, namely the MacWilliams Identities and the Extension Theorem.
	
	A central problem in coding theory is constructing linear codes with optimal parameters, i.e. codes that ensure the reliable transmission of data with minimum constraints and efficient encoding and decoding schemes. Given the sheer amount of linear codes, an exhaustive search for optimal ones is almost always practically infeasible. We can work around this problem by limiting our search to a subclass of linear codes with additional algebraic properties. One of such subclasses is that of polycyclic codes, a generalization of the well-known class of cyclic and constacyclic codes, and in which a code can be seen as an ideal of the polynomial ring $R[x]$ modulo an associated polynomial $f\in R[x].$ Polycyclic codes were already known since the 70's. Still, despite that they were not as extensively studied as cyclic codes, until recently when they started getting more attention (see for example \cite{EdgarMaryamSteve,EdgarMaryamSteve2,Sole2,fotue2020polycyclic,sole}). In particular, the special case of polycyclic codes associated with a trinomial over a finite field was considered in \cite{aydin2022polycyclic}, and many properties and interesting uses of these codes were showcased. 
	
	A way to further reduce the task of finding new codes is the introduction of an equivalence relation that preserves the important properties and characteristics of a code. In the context of linear codes endowed with the Hamming distance, monomial equivalence has been widely adopted as the suitable equivalence. This goes back to the early work of MacWilliams, where it was shown that over finite fields, such a relation essentially amounts to having a linear weight preserving isomorphism. Unfortunately, checking whether two linear codes are equivalent is generally computationally challenging \cite{sendrier2013easy}. However, harnessing the additional algebraic structure present in some linear codes can help to overcome this computational limitation. An example of this was proposed in \cite{CHEN20121217} where the authors introduced an equivalence relation ``$\cong_{n}$" called $n$-isometry to classify constacyclic codes of length $n$ over $\mathbb{F}_q.$ For $\lambda$ and $\mu$  in $\mathbb{F}^*_q$ , $\lambda \cong_{n} \mu$ means the existence of an isometry $\phi$ with respect to the Hamming distance between the rings $\mathbb{F}_q[x]/ \langle x^{n} - \lambda \rangle $ and $\mathbb{F}_q[x]/ \langle x^{n} - \mu \rangle $, this induces a biunivocal relation between the set of $\lambda$-constacyclic codes and that of $\mu$-constacyclic codes which conserves the weight distribution as well as the algebraic structure of codes. A limitation of this relation is that it does not always explicitly relate the generator polynomials of corresponding codes. To overcome this, the authors introduced another equivalence relation ``$\sim_{n}$" in \cite{CHEN201460}, called $n$-equivalence. For  $\lambda$ and $\mu$  in $\mathbb{F}^*_q$, $\lambda \sim_{n} \mu$ if the polynomial $\lambda x^n - \mu$ has a root in $\mathbb{F}_q[x]$ . Equivalently, $\lambda \sim_{n} \mu$ means the existence of a nonzero scalar $a$ in $\mathbb{F}_q$ such that the map $\phi_a$ from the ring $\mathbb{F}_q[x]/ \langle x^{n} - \mu \rangle $ to $\mathbb{F}_q[x]/ \langle x^{n} - \lambda \rangle $, defined by $\phi_a(f(x)) = f(ax)$ is an isometry with respect to the Hamming distance. Note that this relation has been used implicitly years before its formal introduction in \cite{CHEN201460}, especially in the case of negacyclic/cyclic equivalence, see \cite{1317117} for example.  An important consequence of these equivalence relations  comes from the fact that given a finite field $\mathbb{F}_q$, the numbers of $n$-isometry and $n$-equivalence classes were given in the aforementioned papers, hence the search of constacyclic codes with good parameters on $\mathbb{F}_q$ is significantly reduced, as the number of constacyclic codes with different parameters cannot exceed that of $n$-isometry classes nor that of $n$-equivalence classes. These two relations have been generalized to the case of constacyclic codes over finite chain rings in \cite{chibloun2024isometry} and skew constacyclic codes over finite fields in \cite{OUAZZOU2025114279}.
	
	In this paper, we discuss polycyclic codes of arbitrary length associated to a trinomial  over a finite chain ring $R,$ and we generalize the study of $n$-isometry and $n$-equivalence to this class of codes.
	
	The rest of the paper is organized as follows: Section \ref{secPrelim} provides a review of the basic
	background on finite chain rings as well as the structure of polycyclic codes over them based on the notion of strong Gr{\"o}bner bases. In Section \ref{sec:tricodes}, we discuss square-free trinomial codes and their relation to repeated root trinomial codes over some finite fields.
	Section \ref{eq} generalizes the relations of $n$-isometry and $n$-equivalence to the case of trinomial codes over finite chain rings, while in Section \ref{sec:Tri1} we discuss the special case of trinomials of the form $x^n-a_1x-a_0\in R[x]$. Finally, in Section \ref{sec:additive} we see how the theory fares in the case of some trinomial additive codes over $R.$
	
	\section{Preliminaries}\label{secPrelim}

	In this section, we will review some topics on finite chain rings and the polynomial description of polycyclic codes over them. The main references on finite chain rings will be \cite{mcdonald1974finite,wan2003lectures} and on polycyclic codes and their description will be \cite{fotue2020polycyclic, NortonSalagean2} and the references therein.
	
	\subsection{Finite chain rings}
	In this paper, all rings will be associative, commutative, and with
	identity. A ring  is called a \emph{local ring} if it has a unique maximal ideal. A local ring is a
	\emph{chain ring} if its lattice of ideals is a chain under inclusion, it is well known that its ideals are principal (see \cite{mcdonald1974finite} for further information). From now on $R$ will denote a finite chain ring and $\mathfrak m$ its maximal ideal unless otherwise stated. Fix a generator $\gamma$ of $\mathfrak m$ (note that it is a nilpotent element), let $s$ be its \emph{nilpotency
		index}, then the chain of ideals of $R$ is given by
	$$ \langle0\rangle= \langle \gamma^s \rangle \subsetneq \langle \gamma^{s-1}\rangle \subsetneq \dots \subsetneq \langle \gamma^1 \rangle = \mathfrak m \subsetneq \langle \gamma^0\rangle = R.$$
	One has  that
	if $1\leq i<j\leq s$ and $\gamma^ic \in \gamma^jR,$ then $c\in \gamma^{j-i}R.$ In particular, if $\gamma^jR=0$ then $c\in \gamma^{s-j}R$, see   \cite[Corollary 2.3]{NortonSalagean2}.
	
	We will denote the natural map on the residue field by $\bar\dot: R\rightarrow R/\mathfrak m$. If $f(u)$ is a polynomial in $R[u]$ we can extend the previous mapping to define $\bar f(u)\in \left( R/\mathfrak m\right)  [u]$ where $\bar\dot$ is applied to each of the coefficients of $f(u)$. A particular example of chain rings are \emph{Galois rings} that are defined as $\mathrm{GR}(p^{m},r)=\mathbb Z_{p^m}[u]/\langle f(u)\rangle$, where $f(u)$ is a \emph{basic irreducible} polynomial of degree $r$ over $\mathbb{Z}_{p^m},$ i.e. $f(u)\in \mathbb{Z}_{p^m}[x] $ such that $\bar f(u)$ is an irreducible polynomial. In the Galois ring $\mathrm{GR}(p^{m},r)$ there exists a nonzero element $\omega$ of multiplicative order $p^r-1$ which is a root of a basic irreducible polynomial of degree $r$ over $\mathbb{Z}_{p^m}$ and dividing $x^{p^r-1}-1$ in $\mathbb{Z}_{p^m}[x]$ and $\mathrm{GR}(p^{m},r)=\mathbb{Z}_{p^m}[\omega].$ The set $T=\{0,1,\omega,\omega^2,...,\omega^{p^r-2}\}$ is called the \emph{Teichm\"uller set} of the Galois ring $GR(p^m , r).$ Galois rings play a central role in describing chain rings due to the following result.
	\begin{theorem}(\cite[Theorem XVII.5]{mcdonald1974finite})\label{fromFiniteChain}
		
		A finite chain ring $R$ with invariants $(p,m,r,e,t)$, is of the form \begin{equation}\label{fcrAsGrQuotient}
			R \simeq \frac{\mathrm{GR}(p^{m},r)[u]}{\langle g(u),p^{m-1}u^{t}\rangle}
		\end{equation}
		where $g(u) \in \mathrm{GR}(p^{m},r)[u]$ is an Eisenstein polynomial of degree $e$, i.e. $g(u)=u^{e}-p(a_{e-1}u^{e-1}+\ldots+a_{0})$, with $a_{0}\in \mathrm{GR}(p^{m},r)^{\times}$, the set of units of $\mathrm{GR}(p^{m},r)$.
	\end{theorem}
	The maximal ideal of $R$ when expressed in the form of (\ref{fcrAsGrQuotient}) is $\mathfrak m=uR,$ and its nilpotency index is $s=(m-1)e-t.$ The unit group $R^{\times}$ of $R$ is of order $p^{r(s-1)}(p^{r}-1)$, and contains a unique cyclic subgroup $T^{*}$ of order $p^{r}-1$. The set $T=T^{*}\cup\{0\}$ is called the \emph{Teichm\"uller set} of $R$, it is the same as the Teichm\"uller set of $\mathrm{GR}(p^{m},r)$ and every element $y$ of $R$ has a unique expression as $y=\xi_{0}+u\xi_{1} + \ldots + u^{s-1}\xi_{s-1}$   with $\xi_{i}\in T$ for all $i$. That is 
	\begin{equation}\label{eqGeneralFormUnitGroup}
		R^{\times}=T^{*}.\left(1+uR\right)\simeq T^{*}\times(1+uR)
	\end{equation}
	
	Given a polynomial $g(x)\in R[x]$,  we say that $g(x)$ is \emph{a regular polynomial} if it is not a zero-divisor. A regular polynomial $g(x)$ is \emph{ basic irreducible} if $\bar g(x)$ is irreducible in  $\left( R/\mathfrak m\right)  [x]$.   In general, given a finite chain ring $R$ the polynomial ring $R[x]$ is not a unique factorization ring, but we have the following factorization theorem.
	\begin{theorem}\label{hensel}(Hensel lifting, \cite[Theorem XIII.4]{mcdonald1974finite})
		Let $g(x) \in R[x]$ be a monic polynomial. Since $R/\mathfrak m$ is a field,  there are monic, pairwise coprime polynomials $f_1 , . . . , f_k \in \left(R/\mathfrak m\right)[x]$ such that $ \bar{g}=\prod^k_{i=1} f_i$. 	Then there are monic, pairwise coprime polynomials $g_1 , . . . , g_k \in R[x]$ such that $g=\prod^k_{i=1} g_i$	and $\bar{g_i} = f_i$ for $1 \leq i \leq k$.
	\end{theorem}
	Whereas Hensel's lemma does not generally guarantee that the factorization is unique, the following result ensures it for a special case. Note that a polynomial $f(x)\in \left( R/\mathfrak m\right)  [x]$ is called  \emph{square-free} if, for each $g(x)$ in $\in \left( R/\mathfrak m\right)  [x]$, $g^2 \mid f$ implies that $g$ is a unit.
	\begin{proposition}(\cite[Theorem 2.7]{NortonSalagean2})\label{2.7norton}
		If $g \in R[x]$ is a monic polynomial  and $\bar{g}\in \left( R/\mathfrak m\right)  [x]$ is square-free, then $g$ factors uniquely into monic, coprime basic irreducible polynomials.
	\end{proposition}
	Although $R[x]$ is not an Euclidean domain, we have the following version of the Euclidean algorithm for polynomials over finite chain rings.
	\begin{proposition}(\cite[Proposition 2.8]{1317117})\label{divAlg}
		Let $f(x),g(x)$ be two nonzero polynomials in $R[x].$ If $g(x)$ is regular, then there exist polynomials $q(x),r(x) \in R[x]$ such that $f(x)=q(x)g(x)+r(x)$ and $\mathrm{deg}(r)<\mathrm{deg}(g).$ 	  	
	\end{proposition}
	
	\subsection{Strong Gr{\"o}bner bases over finite chain rings}
	
	The notions of \emph{Gr{\"o}bner bases} and \emph{Strong Gr{\"o}bner bases} (SGB) over rings have been well studied in the past decades. For a thorough treatment of these notions see  for example \cite{adams2022introduction,NORTON2003237,norton2001strong}. Over finite chain rings, these notions coincide as it was proven in \cite[Proposition 3.9]{norton2001strong}. We refer to that paper for a definition of Gr\"obner basis over rings and the strong reduction associated to them. A Gr{\"o}bner basis $G$ for an ideal $I$ is called \textit{minimal} if no proper subset of $G$ is a Gr{\"o}bner basis for $I$. On a finite chain ring a \textit{minimal strong Gr{\"o}bner basis} is simply a minimal Gr{\"o}bner basis. Given a finite chain ring, each non-zero ideal of the polynomial ring has a minimal SGB.   We will use the following characterization.
	
	\begin{theorem}(\cite[Theorem 3.2 and Theorem 3.4]{NORTON2003237})\label{3.2NORTON2003237}
		Let $R$ be a finite chain ring and $I\subset R[x]$ an ideal of $R[x]$.  $G \subset R[x]$  is  a minimal strong Gr{\"o}bner basis of $I$ if and only if  $\langle G\rangle = I,$ and 
		\begin{itemize}
			\item[(i)] for some $1\leq u\leq s-1,$ $G= \{\gamma^{\lambda_0}g_0,\gamma^{\lambda_1}g_1,\ldots,\gamma^{\lambda_u}g_u\}$,  where $0\leq \lambda_0 < \lambda_1 <\ldots<\lambda_u \leq s-1$ and $g_i\in R[x]$ for all $0\leq i\leq u.$
			\item[(ii)] for $0\leq i\leq u,$ the leading coefficient of $g_i$ is a unit in $R.$
			\item[(iii)] $\mathrm{deg}(g_0) > \mathrm{deg}(g_1) > \ldots> \mathrm{deg}(g_u).$
			\item[(iv)] for $0\leq i\leq u-1,$ $\gamma^{\lambda_{i+1}}g_i \in \langle \gamma^{\lambda_{i+1}}g_{i+1},\ldots,\gamma^{\lambda_u}g_u \rangle$ in $R[x].$
		\end{itemize}
		Moreover, if there
		is a monic $f\in I$ such that $\bar{f}\in \left( R/\mathfrak m\right)  [x]$ is square-free, then $I$ has a minimal SGB $G=\{\gamma^{\lambda_0}g_0,\gamma^{\lambda_1}g_1,\ldots,\gamma^{\lambda_u}g_u\}$ which satisfies \textit{(i)-(iii)} above, with the additional property 
		\begin{itemize}
			\item[(iv')] $g_{u} | g_{{u-1}} | \ldots | g_{0} |f$ and $\lambda_0=0.$
		\end{itemize}
		
	\end{theorem}
	
	\subsection{Polycyclic codes over finite chain rings}
	A \emph{polycyclic code} $C$ over the ring $R$ can be described as an ideal of the ring $R[x]/\langle f(x)\rangle$ where $f$ is a monic polynomial in $R[x]$. They were introduced formally in \cite{SergioSteve} and they are a generalization of cyclic and constacyclic codes which have been extensively studied in coding theory literature. Polycyclic codes over finite fields have been studied from several points of view, see for example  \cite{sole,Sole2,Hassan} and the references therein. They have been also studied over chain rings and local rings and also in the multivariable case, for example in \cite{fotue2020polycyclic,EdgarMaryamSteve,EdgarMaryamSteve2, EdgarInaki,EdgarInaki2}. 
	Given an ideal $I$ of $R[x]$, we will denote as $q_I$ the canonical projection $q_I:R[x] \longrightarrow R[x]/I,$ and $q_f$ when $I=\langle f\rangle$. When there is no ambiguity, we will denote the projection simply $q$. It is well-known that there is a one-to-one order-preserving correspondence between the ideals of $R[x]/I$ and the ideals of $R[x]$ that contain $I$, such that for each ideal $I\subset J\subset R[x] $ it corresponds to $q_I(J)$.  
	The following result is a generalization of   \cite[Theorem 4.2]{NORTON2003237} to the case of polycyclic codes. Remark that we do not impose on the polynomial defining the ideal to have simple roots.
	
	\begin{proposition}\label{genSetExist}
		Let $f(x)\in R[x]$ be a monic polynomial and let $C \subset R[x]/\langle f(x)\rangle$ be a non-zero ideal. There is a positive integer $u\leq s-1$ and a set  $G=\{\gamma^{\lambda_0}g_0,\gamma^{\lambda_1}g_1,\ldots,\gamma^{\lambda_u}g_u\} \subset R[x]$ such that $C= \langle q(G)\rangle$ and 
		\begin{itemize}
			\item[(i)] $g_0,\ldots,g_u$ are monic polynomials in $R[x]$;
			\item[(ii)] $0\leq \lambda_0 < \lambda_1 <\ldots<\lambda_u \leq s-1$;
			\item[(iii)] $\mathrm{deg}(f)>\mathrm{deg}(g_0) > \mathrm{deg}(g_1) > \ldots> \mathrm{deg}(g_u)$;
			\item[(iv)] $\gamma^{\lambda_{i+1}}g_i \in \langle \gamma^{\lambda_{i+1}}g_{i+1},\ldots,\gamma^{\lambda_u}g_u \rangle \subset R[x]$ for $0\leq i\leq u-1,$ and $\gamma^{\lambda_0}f \in \langle G\rangle.$
		\end{itemize}
	\end{proposition}

	Since $f$ is a monic polynomial, by Proposition~\ref{divAlg} each residue class in $R[x]/\langle f\rangle$ has a unique representative of degree less strictly than the degree of $f$. Hence, as property \textit{(iii)} in Proposition~\ref{genSetExist} states that every polynomial of $G$ has a degree strictly less than $\mathrm{deg}(f)$, we get that $q(G)=G$ and $C=\langle G\rangle.$

	\begin{definition}[Generating Set in Standard Form]
		Let $f(x)\in R[x]$ be a monic polynomial. For $C \subset R[x]/\langle f(x)\rangle$ a non-zero polycyclic code, a set $G=\{\gamma^{\lambda_0}g_0,\gamma^{\lambda_1}g_1,\ldots,\gamma^{\lambda_u}g_u\} \subset R[x]$ such that $C= \langle q(G)\rangle$  will be called a generating set in standard form of the code $C$ if it fulfills the conditions (i), (ii), (iii) and (iv) in Proposition~\ref{genSetExist}.
	\end{definition}

	\begin{lemma}(\cite[Proposition 4.1]{NORTON2003237})\label{prop41NORTON2003237}
		Let $I$ be an ideal of $R[x]$ containing a monic polynomial $f$ and let $G$ be an SGB for $I.$ Then for $g\in R[x],$ $q_f(g)\in q_f(I)$ if and only if $g$ strongly reduces to $0$ with respect to $G.$
	\end{lemma}

	\begin{lemma}\label{genSetIfMonicIn}
		Let $I \subset R[x]$ be a nonzero ideal that contains a monic polynomial $f,$ and let $G=\{\gamma^{\lambda_0}g_0,\gamma^{\lambda_1}g_1,\ldots,\gamma^{\lambda_u}g_u\}$ be a minimal SGB of $I$ such that $\lambda_0<\lambda_1<\ldots<\lambda_u.$ Then we have $\lambda_0=0$ and $\mathrm{deg}(g_0)\leq \mathrm{deg}(f).$
	\end{lemma}
	\begin{proof}
		As the monic polynomial $f\in I=\langle G\rangle$, then we have $\langle G\rangle \nsubseteq \gamma R[x],$ hence $\lambda_0 =0.$ Moreover, suppose that  $\mathrm{deg}(g_0)> \mathrm{deg}(f).$ There exist polynomials $\alpha_0,\ldots,\alpha_u \in R[x]$ such that $ f=\alpha_0g_0 + \sum_{i=1}^{u}\gamma^{\lambda_i}\alpha_ig_i=\alpha_0g_0 + \gamma^{\lambda_1}\sum_{i=1}^{u}\gamma^{\lambda_i-\lambda_1}\alpha_ig_i.$ Thus, as $f$ is monic, its leading monomial must be that of $\alpha_0g_0,$ and as $g_0$ is also monic it follows that $\mathrm{deg}(g_0)\leq \mathrm{deg}(f).$
	\end{proof}
	
	\begin{proposition}\label{genSetMinSGB}
		Let $f(x)\in R[x]$ be a monic polynomial and let $I \subset R[x]$ be a non-zero ideal that contains $f$. We have the following 
		\begin{itemize}
			\item[(i)] If $G=\{\gamma^{\lambda_0}g_0,\gamma^{\lambda_1}g_1,\ldots,\gamma^{\lambda_u}g_u\}$ is a generating set in standard form of $q(I)\subset R[x]/\langle f(x)\rangle,$ where $\lambda_0<\lambda_1<\ldots<\lambda_u,$ then either $\lambda_0 = 0$ and $G$ is a minimal SGB of $I,$ or $\lambda_0 \neq 0$ and $\{f\}\cup G$ is a minimal SGB of $I.$
			\item[(ii)]  If $G=\{g_0,\gamma^{\lambda_1}g_1,\ldots,\gamma^{\lambda_u}g_u\}$ is a minimal SGB of $I,$ then either $\mathrm{deg}(g_0)<\mathrm{deg}(f)$ and $G$ is a generating set in standard form of $q(I),$ or $G\backslash\{g_0\}$ is a generating set in standard form of $q(I).$
		\end{itemize}
	\end{proposition}
	\begin{proof} $\quad $
		\begin{itemize}
			\item[(i)] Suppose that $\lambda_0 = 0.$ By construction, a generating set in standard form  $G$ is clearly  a minimal SGB, so all we need to prove is that $\langle G\rangle=I.$ As $f\in I$ and $\langle q(G)\rangle=q(I),$ it is easy to see that $\langle G\rangle \subset I.$ Let $g$ a polynomial of the ideal $I.$ Using property $(iv)$ of Proposition \ref{genSetExist} we see that $f\in \langle G\rangle,$ hence using Lemma \ref{prop41NORTON2003237}, $q(g)\in q(I)=\langle q(G)\rangle= q(\langle G\rangle)$ implies that $g\in \langle G\rangle$ and thus $I=\langle G\rangle.$ If $\lambda_0 \neq 0,$ then we can easily verify that $\{f\}\cup G$ is a minimal SGB contained in $I.$ Thus, using the same argument, we find that $\{f\}\cup G$ is a minimal SGB of $I.$
			
			\item[(ii)] Suppose that $\mathrm{deg}(g_0)<\mathrm{deg}(f).$ As $\langle G\rangle=I,$ $q(I)=q(\langle G\rangle) =\langle q(G)\rangle$ and it is easy to check that  $G$ verifies all the properties of Proposition \ref{genSetExist}, hence it is a generating set in standard form of $q(I)$. 	
			Otherwise, if $\mathrm{deg}(g_0)\geq \mathrm{deg}(f),$ then by Lemma \ref{genSetIfMonicIn}, $\mathrm{deg}(g_0)=\mathrm{deg}(f).$ As both $f$ and $g_0$ are monic, there exists a polynomial $r \in R[x]$ such that $g_0=f+r$ and $\mathrm{deg}(r)< \mathrm{deg}(f).$ Moreover, as $g_0$ and $f$ are both in $I,$  $r \in I=\langle g_0,\gamma^{\lambda_1}g_1,\ldots,\gamma^{\lambda_u}g_u\rangle.$ Let us prove that $r \in \langle\gamma^{\lambda_1}g_1,\ldots,\gamma^{\lambda_u}g_u\rangle$. Let $\alpha_0,\ldots,\alpha_u$ be polynomials in $R[x]$ such that $r=\alpha_0g_0+\sum_{i=1}^{u}\alpha_i\gamma^{\lambda_i}g_i.$ If $\alpha_0g_0=0$ then we are done. Else, as $g_0$ is a monic polynomial we have $\mathrm{deg}(r) <\mathrm{deg}(g_0)\leq \mathrm{deg}(\alpha_0g_0),$ so for some $1\leq i\leq u,$ we must have $\alpha_i\gamma^{\lambda_i}g_i\neq 0.$		
			To simplify the notations, \textit{w.l.o.g} we take  $\alpha_1\gamma^{\lambda_1}g_1\neq 0.$ 
			As $\mathrm{deg}(r)< \mathrm{deg}(\alpha_0g_0),$ we must have that each term of  $\alpha_0g_0$ of degree greater than that of $r$ must vanish in the sum $\alpha_0g_0+\sum_{i=1}^{u}\alpha_i\gamma^{\lambda_i}g_i.$ But as all terms of the polynomials  $\alpha_1\gamma^{\lambda_1}g_1,\ldots,\alpha_u\gamma^{\lambda_u}g_u$ have a coefficient of the form $\gamma^{\lambda_1}\delta$ for some $\delta \in R,$ the leading coefficient of $\alpha_0$ must be in $\gamma^{\lambda_1}R$ and it follows by induction that the polynomial $\alpha_0$ is of of the from $\gamma^{\lambda_1}\alpha_0^*$ for some $\alpha_0^*\in R[x].$ Hence $\alpha_0g_0=\alpha_0^*\gamma^{\lambda_1}g_0 \in \langle\gamma^{\lambda_1}g_1,\ldots,\gamma^{\lambda_u}g_u\rangle$ from property $(iv)$ of a minimal SGB. Thus we have $r \in \langle\gamma^{\lambda_1}g_1,\ldots,\gamma^{\lambda_u}g_u\rangle$ and it follows that $q(I)=\langle q(G)\rangle=\langle q(g_0),\gamma^{\lambda_1}q(g_1),\ldots,\gamma^{\lambda_u}q(g_u)\rangle=\langle \gamma^{\lambda_1}q(g_1),\ldots,\gamma^{\lambda_u}q(g_u)\rangle=
			\langle q(G\backslash \{g_0\})\rangle.$ All we have left to show is that $\gamma^{\lambda_1}f \in \langle G\backslash \{g_0\}\rangle,$ but that is easy to see from the fact that  $f=g_0-r,$  $\gamma^{\lambda_1}g_0 \in \langle G\backslash \{g_0\}\rangle$ and $r \in \langle G\backslash \{g_0\}\rangle.$
		\end{itemize}
	\end{proof}
	
	From  Proposition~\ref{genSetMinSGB}, for a polycyclic code $C,$ a generating set in standard form can be obtained from a minimal SGB of the corresponding ideal of $C$ in $R[x]$ by simply omitting those polynomials of degree greater or equal to $\mathrm{deg}(f).$
	Thus, it is clear that for a polycyclic code $C$ with generating set in standard form $\{\gamma^{\lambda_0}g_0,\ldots,\gamma^{\lambda_u}g_u\},$ $u$ and the $(u+1)$-tuples $(\lambda_0,\ldots,\lambda_u)$ and $(\mathrm{deg}(g_0),\ldots,\mathrm{deg}(g_u))$ are invariants of the code. 
	
	In the case where $f(x)\in R[x]$ is a square-free polynomial, the previous characterization is sharper as proved in \cite[Theorem 4.4, Corollary 4.6, Remark 4.7]{NortonSalagean2} that provides the following result.

	\begin{proposition}(\cite{NortonSalagean2})\label{polyPrincipal}
		Let $f(x)\in R[x]$ be a monic polynomial such that $\overline{f}\in \left( R/\mathfrak m\right)[x]$ is square-free. 
		\begin{enumerate}
			\item Each non-zero polycyclic code $C \subset R[x]/\langle f(x)\rangle$ has a generating set in standard form $\{\gamma^{\lambda_0}g_0,\gamma^{\lambda_1}g_1,\ldots,\gamma^{\lambda_u}g_u\}$ such that $g_{u} | g_{{u-1}} | \ldots | g_{0} |f.$ Furthermore, such a generating set is unique, we will call it the generating set in the standard form of $C.$ 
			\item $R[x]/\langle f(x)\rangle$ is a principal ideal ring and for each ideal $C$ one has that $$C=\left\langle\sum_{i=0}^{u}\gamma^{\lambda_i}g_i\right\rangle,$$
			where $\{\gamma^{\lambda_0}g_0,\gamma^{\lambda_1}g_1,\ldots,\gamma^{\lambda_u}g_u\}$ is the generating set in the standard form of $C.$ 
		\end{enumerate}
	\end{proposition}

	\section{Trinomial codes}\label{sec:tricodes}
	Given a chain ring $R$, we will focus on polycyclic codes associated with monic trinomials of the
	form $x^n-ax^i- b\in R[x]$ where $n > i > 0$ and $a, b\in  R ^\times$. From now on,  whenever we refer to a
	trinomial those conditions will hold, and by \emph{trinomial code}
	we mean a polycyclic code associated with a trinomial. In \cite{aydin2022polycyclic}, the authors obtained many results for polycyclic codes over finite fields associated with a trinomial
	from the factorization of the trinomial.  Note that for a polynomial in $\mathbb{F}_q[x]$  its \emph{discriminant} $\Delta(f)$ is nonzero  if the polynomial $f$ is square-free. The following result provides a characterization of the discriminant in the case of trinomials.
	
	\begin{proposition}(\cite[Theorem 4]{GREENFIELD1984105})
		The trinomial $f(x) = x^n + ax^k + b\in \mathbb{F}_q[x]$ has discriminant \begin{equation}
			\Delta (f)=(-1)^{n(n-1)/2}b^{(k-1)}\left[n^Nb^{N-K}-(-1)^N(n-k)^{N-K}k^Ka^N \right] ^d,
		\end{equation}
		where $d=\mathrm{gcd}(n,k)$, and $N=n/d, ~ K=k/d.$
	\end{proposition}
	
	In the case of trinomials defined over chain rings, we have the following criteria for checking if it is square-free.
	
	\begin{lemma} \label{sqFrCriter} Let $f(x)=x^n+ax^k+b,$ be a trinomial in $R[x]$ where $R$ is a chain ring with residue field of characteristic $p$.
		\begin{enumerate}
			\item If  $p$ divides $nk$ but not $n+k$, then $f$ has a square-free projection in $\left( R/\mathfrak m\right)  [x]$.
			\item In particular, for all $n\in p\mathbb{Z}\cap\mathbb{N}^*$ and $a,b\in R^\times,$ $\overline{x^{n}+ax+b}$ and $\overline{x^{n}+ax^{n-1}+b}$ are square-free trinomials in $\left( R/\mathfrak m\right)  [x]$.
		\end{enumerate}
	\end{lemma}
	\begin{proof}
		As $p$ divides $nk$ and not $n+k$ it must divide either $n$ or $k$ but not both. Hence, using the notations of the previous Proposition, either $p$ divides $n^N$ or it divides  $(n-k)^{N-K}k^K$. Thus, the discriminant of $\overline{x^n+ax^k+b}\in  \left( R/\mathfrak m\right)  [x]=\mathbb{F}_{p^r}[x],$ is of one of  these forms $$\displaystyle (-1)^{n(n-1)/2}\overline{b}^{(k-1)}(\alpha \overline{b}^{N-K}) ^d\hbox{  or  } (-1)^{n(n-1)/2}\overline{b}^{(k-1)}(\beta \overline{a}^N ) ^d,$$ where $\alpha, \beta \in \mathbb{F}_{p^r}^*. $ Hence as $a,b\in R^\times,$ the discriminant of $\overline{x^n+ax^k+b}$ is always nonzero, i.e.  $\overline{x^n+ax^k+b}$ is square-free.
	\end{proof}

	\begin{corollary}
		Let $n,k \in \mathbb{N}^*$ such that $p$ divides $nk$ but not $n+k.$ For all $a,b\in R^\times,$ the ring $R[x]/\langle x^n+ax^k+b\rangle$ is principal.
	\end{corollary}
	\begin{proof}
		Follows immediately from Proposition \ref{polyPrincipal} and Lemma \ref{sqFrCriter}.
	\end{proof}
	
	We will consider now  some repeated-root polynomials  over the finite field $\mathbb F_{p^r}$. The following lines could be found more detailed in \cite[Section 5]{EdgarMaryamSteve2}.  Let $f(x)\in \mathbb F_{p^r}[x]$ a simple-root polynomial of degree $n$ and order $e,$ and let  $f (x) =\prod_{i=1}^{\nu}f_ i (x)$ be its unique factorization. It is clear that $f (x^{p^k}) =\prod_{i=1}^{\nu} f_ i (x^{p^k})$   for $k>0$
	and  for each $1 \leq i \leq \nu$, there exists an irreducible polynomial
	$g_i (x)$ in $\mathbb F_{p^r}[x]$ such 
	that $f_i (x^{p^k} ) = g_i (x)^{p^k}$. From now on, we will assume that $R$ is the ring 
	\begin{equation}
		R=\mathbb F_{p^r}[x]/\langle f(x^{p^k})\rangle =  \mathbb F_{p^r}[x]/\left\langle \left(\prod_{i=1}^{\nu} g_i(x)\right)^{p^k}\right\rangle
	\end{equation}
	and we will have that  $N = np^k$.
	One can write any element $a(x)\in R$ as $a_0 (x) + a_1(x) x^{p^k} +\ldots + a_{n-1}(x) x^{(n-1)p^k}$, where $a_i(x)\in {\mathbb{F}}_{p^r}[x]$.
	Let $S$ be the ring $\mathbb F_{p^r}[x,y]/\langle x^{p^k} -y, f (y)\rangle$. We
	have that any ideal of the ring $R$ is principally generated by a divisor of $f(x^{p^k} )$.
	In fact, it is of the form   $\langle G(x)\rangle$, where $G(x)=\prod_{j=1}^\nu g_i(x)^{i_j}$
	and $0 \leq i_j \leq p^k$.
	The
	map $\varphi:R \rightarrow S$ given by
	$\varphi\left(\sum_{i=0}^{n-1} a_i(x)x^{i p^k}\right)=a(x,y)= \sum_{i=0}^{n-1} a_i(x)y^i$
	is a ring isomorphism.
	Now consider the ring
	\begin{equation}\label{eq:reproots}
		U=\mathbb F_{p^r}[x,y]/\langle{x^{p^k}-1,f(y)}\rangle=\left(\mathbb F_{p^r}[x]/\langle{x^{p^k}}-1\rangle\right)[y]/\langle f(y)\rangle,
	\end{equation}
	and  denote as $W$ the ring $W=\mathbb F_{p^r}[x]/\langle{x^{p^k}-1\rangle}$. Note that  $W$  is a finite chain ring whose maximal ideal is $\langle (x-1)\rangle$. 
	The map
	$\psi:S\rightarrow U$ defined by $\psi(a(x,y))=a(y^{e'}x,y)$ is a ring isomorphism, where $e'$ is the inverse of $p^k$ in $\mathbb Z_e$.
	Thus,  one has that an ideal is a polycyclic code  in $\mathbb F_{p^r}[x]/\langle f(x^{p^k})\rangle $  if and only if
	$\mu(C)=\psi(\varphi(C))$ is a  polycyclic code in  $W[y]/\langle f(y)\rangle$. 
	
	Hence, from the previous discussion, the study of some trinomial codes such that $p$ divides both $n$ and $k$ in the finite field case can be done by considering simple-root trinomial codes over chain rings.  More specifically, those cases are the ones where all the irreducible factors of the considered trinomial are square-free and have the same multiplicity $p^z$ for some integer $z\geq 1$. Note that the result in   \cite[Theorem 3.3]{aydin2022polycyclic} classifies those cases.

	\section{Equivalence of trinomial  codes}\label{eq}
	
	In this section, we generalize the notions of $n$-isometry and $n$-equivalence introduced in \cite{CHEN20121217} and \cite{CHEN201460}, respectively, to the case of trinomial codes over a finite chain ring. We will end the section by computing the number of equivalence classes w.r.t. the new $n$-equivalence relation.

	Let us first fix some notations.
	For $0<k<n$, let $B_k$ denote the set of degree $k$ binomials in $R[x]$ that have coefficients in $R^\times$ and do not vanish at zero. It is straightforward that $B_k=\{b_1x^k+b_0 ~|~ b_1,b_0\in R^\times\}$ and that $B_k$ is in a one-to-one correspondence with $R^\times \times R^\times$. We denote $\star$ the component-wise multiplication on $B_k$, i.e. for $a(x)=a_1x^k+a_0$ and $b(x)=b_1x^k+b_0$ in $B_k$, $a(x)\star b(x)=a_1b_1x^k+a_0b_0$. It is easy to see that for each $0<k<n$, $(B_k,\star)$ is an abelian group with identity element $x^k+x$. Throughout the paper, for an element $b(x)\in B_k$ we denote $b(x)^{-1}$ its inverse with respect to  the operation $\star$, i.e. for $b(x)=b_1x^k+b_0,$ $b(x)^{-1}=b_1^{-1}x^k+b_0^{-1}.$
	
	Let $B$ be the set of all binomials in $R[x]$ of degree less or equal to $n-1$ with coefficients in $R^\times$ and that do not vanish at zero. The family $(B_k)_{0<k<n}$ is a partition of    $\displaystyle B=\underset{{0<k<n}}{\mathlarger{\mathlarger{\sqcup}}}B_k$.

	\subsection{Isometry between trinomial codes }
	
	As usual, we will define the \emph{Hamming weight} of a polynomial as the number of non-zero entries in its support. The \emph{Hamming distance} between two polynomials is just the Hamming weight of their difference.

	\begin{definition}\label{isoDef}
		Let $a(x), b(x)$ be two binomials in $B$, we say that an $R$-algebra isomorphism $$ \phi : {R[x]}/ \langle x^{n} -a(x) \rangle \longrightarrow {R[x]}/ \langle x^{n} - b(x) \rangle $$ is an isometry, if it preserves the Hamming distance $d_H(.)$ on the algebras, i.e.
		\[d_H(\phi(f),\phi(f'))=d_H(f,f'),~~ \forall f,f'\in R[x]/ \langle x^{n} - a(x) \rangle.\] If there is an isometry between $\displaystyle {R[x]}/ \langle x^{n} - a(x) \rangle$ and $\displaystyle {R[x]}/ \langle x^{n} - b(x) \rangle$, we say that $a(x)$ is $n$-isometric to $b(x)$, and denote $a(x) \cong_{n} b(x)$.
	\end{definition}
	
	\begin{proposition}
		The relation $\cong_{n}$ is an equivalence relation on the set of binomials $B$.
	\end{proposition}

	\begin{proof}
		Reflexivity: For all $a(x)$ in $B$, $ \displaystyle id_{{R[x]}/ \langle x^{n} - a(x) \rangle}$ is obviously an isometry, hence $ \cong_{n}$ is reflexive. 
		
		Symmetry: For all $a(x)$ and $b(x)$ in $B$, if $b(x) \cong_{n} a(x)$ then there exists an isometry $\phi$ from $R[x]/ \langle x^{n} - b(x) \rangle$ to $R[x]/ \langle x^{n} - a(x) \rangle$, $\phi$ being an isometry we have that $\phi^{-1}$ is an algebra isomorphism. And for all $f$ in $R[x]/ \langle x^{n} - a(x) \rangle$, as $\phi$ preserves the Hamming weight $\mathrm{w}_H()$, $\mathrm{w}_H(f)=\mathrm{w}_H(\phi(\phi^{-1}(f)))=\mathrm{w}_H(\phi^{-1}(f))$. So $\phi^{-1}$ is an isometry and $a(x) \cong_{n} b(x)$.
		
		Transitivity: Let $a(x)$, $b(x)$ and $c(x)$ in $B$ such that $a(x) \cong_{n} b(x)$ and $b(x) \cong_n c(x)$. Let $\phi: R[x]/ \langle x^{n} - a(x) \rangle \rightarrow R[x]/ \langle x^{n} - b(x) \rangle$ and $\psi : R[x]/ \langle x^{n} - b(x) \rangle \rightarrow R[x]/\langle x^{n} - c(x) \rangle$ be two isometries. It is easy to see that $\psi\circ\phi$ is an isometry from $R[x]/ \langle x^{n} - a(x) \rangle$ to $R[x]/\langle x^{n} - c(x) \rangle$, hence $a(x) \cong_n c(x)$.
		
	\end{proof}
	Given two polycyclic codes $C$ and $C^\prime$, a map $\tau:C\rightarrow C^\prime$ is called a \emph{monomial
		transformation} if $\tau$ is a linear homomorphism and has the form $\tau\left(\sum a_i x^i\right)= \sum r_i a_{\pi(i)} x^i$,
	where $\pi$ is a permutation of the set $\{0, 1,\ldots, n-1\}$   and $r_0,\ldots , r_{n-1}$ are elements in $R^\times$. A monomial transformation is clearly an isomorphism. Two codes   are \emph{monomial equivalent} if there exists a monomial transformation
	between them. 
	Note that   $n$-isometry is related to the notion of monomial equivalence as it is shown in the following result. 
	\begin{proposition}\label{isometryImplyMonoEquiv}
		Let $a(x), b(x) \in B$ such that $a(x) \cong_{n} b(x)$ and let $\phi$ be an associated isometry. Let $C$ be a $(x^n-a(x))$-polycyclic code of length $n$ over $R$. Then $\phi(C)$ is a $(x^n-b(x))$-polycyclic code of length $n$ over $R$ and $\phi(C)$ is monomially equivalent to $C$.
	\end{proposition}
	
	\begin{proof}
		There exists a monomial permutation between two codes over a finite chain ring if and only if there exists a linear Hamming isometry (see \cite{wood1997}). And it is immediate that each isometry as defined in Definition \ref{isoDef}, is an $R$-linear isomorphism that preserves Hamming distance.
	\end{proof}
	
	An important consequence of the precedent proposition is that, if $a(x),b(x)\in B$ are $n$-isometric, then the trinomial codes associated to $(x^n-a(x))$ and those associated to $(x^n-b(x))$ are practically the same. This serves a great deal in reducing  the number of trinomials considered when searching for new codes, as only representatives of the $n$-isometry classes should be examined.  
	
	\begin{proposition}\label{isoForm}
		Let $a(x), b(x) \in B$ such that $a(x) \cong_{n} b(x)$, and let $ \phi : {R[x]}/ \langle x^{n} -a(x) \rangle \rightarrow {R[x]}/ \langle x^{n} - b(x) \rangle $ be an isometry. Then there exists a unit $\alpha \in R^\times$ and an integer $0<j<n$ such that $\phi(x)=\alpha x^j$.		
	\end{proposition}
	\begin{proof}
		As $\phi$ preserves the Hamming weight, there exists an element $\alpha$ in $R\backslash (0)$ and $0< j \leq n-1$ such that $\displaystyle \phi(x)=\alpha x^{j}$ in ${R[x]}/ \langle x^{n} - b(x) \rangle $. Note that $j\neq 0$ because $\phi(\alpha)=\alpha$ and $\phi$ is one-to-one.
		If $\alpha$ is not a unit in $R$, then $\alpha^{s}=0$, $s$ being the nilpotency index of $R$. Thus $\phi(x^{s})=\phi(x)^{s}=\alpha^{s}x^{js}=0.$ But as $\phi$ is one-to-one, this means that $x^{s}=0$ in  $R[x]/ \langle x^{n} - a(x) \rangle,$ i.e. there exists a monic polynomial $g(x)\in R[x]$ such that $x^s=g(x)(x^n-a(x))$ hence,  $x^s=\overline{g(x)}\overline{(x^n-a(x))}$ in $\left( R/\mathfrak m\right)  [x]$, but as the coefficients of $a(x)$ are units in $R,$ $x$ does not divide $\overline{x^n-a(x)}$ in $\left( R/\mathfrak m\right)  [x]$ which is a contradiction and hence $\alpha$ must be in $R^{\times}$.
	\end{proof}
	\begin{remark}
		From the previous proposition, we see that an isometry is a modular composition with a nonzero monomial that vanishes at zero.
	\end{remark}

	\subsection{Equivalence between trinomial codes}
	
	Although the $n$-isometry relationship proved to be of great interest in the case of constacyclic codes over finite fields \cite{CHEN20121217}, finite chain rings \cite{chibloun2024isometry} and even for skew constacyclic codes over finite fields \cite{OUAZZOU2025114279}, it is generally quite hard to deal with for polycyclic codes associated to polynomials with more than two terms. Due to that, plus the fact that it does not relate the generators of equivalent codes in an obvious way, we opt for a special case of this relation, where despite losing generality we avoid the aforementioned difficulties. We discuss this relation in detail in the remainder of this section. 
	
	\begin{definition}\label{def:n}
		Let $a(x),b(x)\in B$. We say that $a(x) \text{ and }b(x)$ are $n$-equivalent and denote $a(x) \sim_n b(x)$, if there exists an element $\alpha$ in $R^\times$ such that 
		\begin{align*}
			\begin{array}{cccc}
				\phi:& {R[x]}/ \langle x^{n} -a(x) \rangle & \longrightarrow & {R[x]}/ \langle x^{n} - b(x) \rangle  \\&&&\\ 
				& f(x) & \longrightarrow & f(\alpha x)
			\end{array} 
		\end{align*}
		is an $R$-algebra isomorphism.
	\end{definition}
	
	\begin{proposition}
		$``\sim_n"$ is an equivalence relation on $B$.
	\end{proposition}
	\begin{proof}
		Reflexivity: For all $a(x)$ in $B$, $ \displaystyle id_{{R[x]}/ \langle x^{n} - a(x) \rangle}$ is clearly an $R$-algebra isomorphism of the desired form, so $``\sim_{n}"$ is reflexive. 
		
		Symmetry: For all $a(x)$ and $b(x)$ in $R^{\times}$, if $a(x) \sim_{n} b(x)$ then there exists an element $\alpha$  such that \begin{align*}
			\begin{array}{cccc}
				\phi:& {R[x]}/ \langle x^{n} -a(x) \rangle & \longrightarrow & {R[x]}/ \langle x^{n} - b(x) \rangle  \\&&&\\ 
				& f(x) & \longrightarrow & f(\alpha x)
			\end{array} 
		\end{align*} is an $R$-algebra isomorphism. It is easy to see that $\phi^{-1}$ is the $R$-algebra isomorphism from  $R[x]/ \langle x^{n} -b(x) \rangle$ to $R[x]/ \langle x^{n} -a(x) \rangle$ and we have for each $f(x)\in R[x]/ \langle x^{n} -b(x) \rangle $, $\phi^{-1}(f(x))=f(\alpha^{-1}x)$, hence $b(x) \sim_{n} a(x)$.
		
		Transitivity: Let $a(x)$, $b(x)$ and $c(x)$ in $R^{\times}$ such that $a(x) \sim_{n} b(x)$ and $b(x) \sim_n c(x)$. And let $\phi: R[x]/ \langle x^{n} - a(x) \rangle \rightarrow R[x]/ \langle x^{n} - b(x) \rangle$ and $\psi : R[x]/ \langle x^{n} - b(x) \rangle \rightarrow R[x]/\langle x^{n} - c(x) \rangle$ the two $R$-algebra isomorphisms defined by $\phi(x)=\alpha x$ and $\psi(x)=\beta x$ for $\alpha, \beta \in R^\times$. It is easy to see that $\psi\circ\phi : x \longrightarrow  \beta\alpha x$ is an isometry from $R[x]/ \langle x^{n} - a(x) \rangle$ to $R[x]/\langle x^{n} - c(x) \rangle$, hence $a(x) \sim_n c(x)$.
	\end{proof}
	\begin{remark}\label{equivIsIso}
		It is easy to see that an $R$-algebra isomorphism $\phi$ as defined above is in fact an isometry with respect to the Hamming distance. Thus,  we get the immediate following proposition.
	\end{remark}
	\begin{proposition}\label{EqImpliesIso}
		Let $a(x)$ and $b(x)$ in $B$. If $a(x)$ is $n$-equivalent to $b(x)$ then $a(x)$ and $b(x)$ are $n$-isometric.
	\end{proposition}
	
	The following result relates the generator sets in standard form of polycyclic codes associated with $n$-equivalent binomials.
	\begin{proposition}
		Let $a(x),b(x)\in B$ such that $a(x)\sim_n b(x)$ and $\alpha\in R^\times$ such that $\phi : x \longrightarrow \alpha x$ is an associated isometry. Let $C$ be a $(x^n-a(x))$-polycyclic code of length $n$ over $R$, and let $G= \{\gamma^{\lambda_0}g_0(x),\ldots,\gamma^{\lambda_u}g_u(x)\}$ be a generator set in standard form of $C$. Then $\phi(C)$ is a $(x^n-b(x))$-polycyclic code of length $n$ over $R$, and $G'= \{\gamma^{\lambda_0}\alpha^{-\mathrm{deg}(g_0)}g_0(\alpha x),\ldots,\gamma^{\lambda_u}\alpha^{-\mathrm{deg}(g_u)}g_u(\alpha x)\}$ is a generator set in standard form of $\phi(C)$.
	\end{proposition}
	\begin{proof}
		Given the fact that $\phi$ is an $R$-algebra isomorphism between $R[x]/\langle x^n -a(x)\rangle$ and $R[x]/\langle x^n -b(x)\rangle,$ it is immediate to see that $\phi(C)$ is  a $(x^n-b(x))$-polycyclic code of length $n$ over $R$ generated by $G',$ and that $G'$ verifies the properties of a generating set in standard form.
	\end{proof}

	\begin{proposition}\label{equiv_stable}
		Let $a(x),b(x)\in B$. If $a(x)$ and $b(x)$ are $n$-equivalent, then there exists an integer $0<k<n$, such that $a(x),b(x)\in B_k$.		
		
	\end{proposition}
	\begin{proof}
		Let $a(x)=a_1x^i+a_0$ and $b(x)=b_1x^j+b_0$, for some integers $0<i,j<n$ and $a_1,a_0,b_1,b_0$ units in $ R^\times$. If $a(x) \sim_{n} b(x)$, let $\phi: R[x]/ \langle x^{n} - a(x) \rangle \rightarrow R[x]/ \langle x^{n} - b(x) \rangle$ be an $R$-algebra isomorphism with associated coefficient $\alpha$.
		We have 
		\begin{equation*}
			\phi(a(x))=a(\alpha x)= a_1\alpha^ix^i+a_0 \mod(x^{n} - b(x))
		\end{equation*}
		and as $a(x)=x^n$ in $R[x]/ \langle x^{n} - a(x) \rangle$, we have 
		\begin{equation*}
			\phi(a(x))=\phi(x^n)=\phi(x)^n=\alpha^n x^n=  \alpha^nb_1x^j+\alpha^nb_0 \mod(x^{n} - b(x))
		\end{equation*}
		hence as $\alpha, a_1$ and $b_1$ are units in $R$, and as $i,j<n$, by the uniqueness  of the representation of classes in $R[x]/ \langle x^{n} - b(x) \rangle$ as polynomials of degree strictly less than $n$, we must have $a_1\alpha^ix^i+a_0=\alpha^nb_1x^j+\alpha^nb_0$ in $R[x]$, hence $i=j$.
	\end{proof}
	\begin{corollary}\label{equivIsLocal}
		For each $0<k<n$, the relation $``\sim_n"$ induces an equivalence relation on $B_k$ that we will denote the same. Moreover, we have
		\begin{equation*}
			B/\sim_n ~= \underset{{0<k<n}}{\mathlarger{\mathlarger{\sqcup}}}( B_k/\sim_n)
		\end{equation*} 
		and the number of classes of equivalence of $``\sim_n"$ over $B$ equals the sum of the numbers of classes over each $B_k$.
	\end{corollary}
	\begin{proof}
		The corollary follows easily from Proposition \ref{equiv_stable}, and the fact that for $0<i,j<n$, $i\neq j \Rightarrow B_i \cap B_j = \emptyset$.
	\end{proof}
	
	Now that we have established the local nature of the $n$-equivalence relation, we can focus our study on binomials with the same degree. The next theorem gives equivalent characterizations of $n$-equivalence over $B_k$ for each $0<k<n$.
	\begin{theorem}\label{equivCharac}
		Let $0<k<n$ , and let $a(x)=a_kx^k+a_0$ and $b(x)=b_kx^k+b_0$ two binomials in $B_k$. The following are equivalent
		\begin{itemize}
			\item[(i)] $a(x) \sim_{n} b(x)$.
			\item[(ii)]  There exists an element $\alpha$ in $R^\times$ such that 
			
			\begin{equation*}
				b_k\alpha^{n-k}-a_k=b_0\alpha^n-a_0=0
			\end{equation*}
			\item[(iii)] $a(x)\star b(x)^{-1} \in H_k=\{\alpha^{n-k}x^k + \alpha^n ~|~ \alpha \in R^\times\}$.
		\end{itemize}
	\end{theorem}
	\begin{proof}$\quad $
		\begin{description}
			\item[$(i)\Rightarrow(ii)$] $a(x) \sim_{n} b(x)$ means that there exists $\alpha$ in $R^\times$ such that \begin{align*}
				\begin{array}{cccc}
					\phi:& {R[x]}/ \langle x^{n} -a(x) \rangle & \longrightarrow & {R[x]}/ \langle x^{n} - b(x) \rangle  \\&&&\\ 
					& f(x) & \longrightarrow & f(\alpha x)
				\end{array} 
			\end{align*}
			is an $R$-algebra isomorphism. From the proof of Proposition \ref{equiv_stable} we can see that $a_k\alpha^k=b_k\alpha^n$ and $a_0=b_0\alpha^n$, hence $b_k\alpha^{n-k}-a_k=b_0\alpha^n-a_0=0.$
			\item[$(ii)\Rightarrow(iii)$] Suppose there exists $\alpha$ in $R^\times$ such that $b_k\alpha^{n-k}-a_k=b_0\alpha^n-a_0=0.$ We have 
			\begin{align*}
				a(x)\star b(x)^{-1} = a_kb_k^{-1}x^k+a_0b_0^{-1} 			
				=\alpha^{n-k}x^k+\alpha^{n} \in H_k
			\end{align*}
			\item[$(iii)\Rightarrow(i)$] Suppose $a(x)\star b(x)^{-1} \in H_k$, i.e. there exists $\alpha$ in $R^\times$ such that $a_kb_k^{-1}x^k+a_0b_0^{-1} 			
			=\alpha^{n-k}x^k+\alpha^{n}.$ Consider the map 
			\begin{align*}
				\begin{array}{cccc}
					\tilde{\phi}:& {R[x]} & \longrightarrow & {R[x]}/ \langle x^{n} - b(x) \rangle  \\&&&\\ 
					& f(x) & \longrightarrow & f(\alpha x)
				\end{array} 
			\end{align*}
			$\tilde{\phi}$ is a surjective $R$-algebra homomorphism. Indeed, $\tilde{\phi}$ is obviously an $R$-algebra homomorphism, and the surjectivity follows from the fact that $x=\tilde{\phi}(\alpha^{-1}x)$ in ${R[x]}/ \langle x^{n} - b(x) \rangle .$  We have 
			\begin{align*}
				\tilde{\phi}(x^n-a(x))=&\alpha^nx^n-a(\alpha x)=\alpha^n x^n - a_k\alpha^k x^k - a_0 = \\ =& \alpha^n( x^n - a_k\alpha^{k-n} x^k - \alpha^{-n}a_0)=\alpha^n( x^n - b_k x^k - b_0) = \\ =& 0 \mod(x^{n} - b(x))
			\end{align*}
			hence $\langle x^{n} - a(x)\rangle \subseteq \mathrm{ker}(\tilde{\phi}).$ On the other hand, for $f(x)\in \mathrm{ker}(\tilde{\phi})$, there exists a polynomial $g(x)\in R[x]$ such that $\tilde{\phi}(f(x))=f(\alpha x)=g(x)(x^{n} - b(x))$, hence 
			\begin{align*}
				f(x)=& g(\alpha^{-1}x)(\alpha^{-n}x^{n} - b(\alpha^{-1}x))=g(\alpha^{-1}x)(\alpha^{-n}x^{n} - b_k\alpha^{-k}x^k-b_0)=\\= &\alpha^{-n}g(\alpha^{-1}x)(x^{n} - b_k\alpha^{n-k}x^k-b_0\alpha^{n}) = \alpha^{-n}g(\alpha^{-1}x)(x^{n} - a_kx^k-a_0) = \\= & \alpha^{-n}g(\alpha^{-1}x)(x^{n} - a(x))
			\end{align*}
			thus we have $\langle x^{n} - a(x)\rangle = \mathrm{ker}(\tilde{\phi}),$ which gives us that \begin{align*}
				\begin{array}{cccc}
					\phi:& {R[x]}/ \langle x^{n} -a(x) \rangle & \longrightarrow & {R[x]}/ \langle x^{n} - b(x) \rangle  \\&&&\\ 
					& f(x) & \longrightarrow & f(ax)
				\end{array} 
			\end{align*}
			is an $R$-algebra isomorphism, i.e. $a(x) \sim_{n} b(x).$
		\end{description}	
	\end{proof}
	\begin{corollary}\label{cardnialClass}
		For $0<k<n$, let $a(x) \in B_k$. The $n$-equivalence class of $a(x)$ is $a(x)\star H_k,$ and its cardinal is the cardinal of the set $H_k=\{\alpha^{n-k}x^k + \alpha^n ~|~ \alpha \in R^\times\}.$ 
	\end{corollary}
	\begin{proof}
		The result follows easily from $(iii)\Leftrightarrow(i)$ of Theorem \ref{equivCharac} and the fact that the coefficients of binomials in $B_k$ are units in $R$.
	\end{proof}
	
	Following the notations of \cite{chibloun2024isometry}, recall that two elements $a,b$ in $R^\times$ are said to be \emph{$n$-equivalent}, if there exists an element $\alpha$ in $R^\times$ such that the morphism $\phi : R[x]/ \langle x^{n} -a \rangle  \rightarrow R[x]/ \langle x^{n} - b \rangle $ where $\phi(x)=\alpha x$ is an $R$-algebra isomorphism. The following corollary relates the relation of $n$-equivalence between trinomial codes, to that of constacyclic codes.
	\begin{corollary}
		Let $0<k<n$ , and let $a(x)=a_kx^k+a_0$ and $b(x)=b_kx^k+b_0$ be two binomials in $B_k$. If $a(x)$ is $n$-equivalent to $b(x)$ then we have $a_k$ is $(n-k)$-equivalent to $b_k$ and $a_0$ is $n$-equivalent to $b_0$. 
	\end{corollary}
	\begin{proof}
		Follows immediately from Theorem \ref{equivCharac} and \cite[Theorem 3.1]{chibloun2024isometry}.
	\end{proof}
	
	\begin{corollary}
		Let $0<k<n$ such that $\gcd(n,k)=1,$ and let $a(x)=a_kx^k+a_0 \in B_k$ such that $a_0^{n-k}=a_k^n.$ Then each trinomial code associated to $x^n-a(x)$ is equivalent to a trinomial code associated to $x^n-x^k-1.$
	\end{corollary}
	\begin{proof}
		As $\gcd(n,k)=1$ we have $\gcd(n,n-k)=1.$ Let $u \text{ and }v$ be two integers such that $un+v(n-k)=1,$ and let $\alpha = a_k^va_0^u.$ We have $\alpha^{n-k}=(a_k^va_0^u)^{n-k}=a_k^{v(n-k)}(a_0^{n-k})^u=a_k^{v(n-k)}a_k^{un}=a_k,$ and $\alpha^n=(a_k^va_0^u)^n=(a_k^n)^va_0^{un}=a_0^{v(n-k)}a_0^{un}=a_0.$ Hence $a(x)\in H_k$ and the result follows from Theorem \ref{equivCharac}. 
	\end{proof}
	
	\begin{remark}\label{nbrEqIsIndex}
		It is easy to see that for each $0<k<n$, $H_k$ is a subgroup of $(B_k,\star).$ In fact it is the normal subgroup associated to the equivalence relation $``\sim_n"$ on $B_k.$ Using this observation, Corollary \ref{cardnialClass} follows immediately, and we can see that the number of $n$-equivalence classes on $B_k$ is the index $[B_k:H_k]$ of the subgroup $H_k$ in $B_k.$
	\end{remark}
	
	\subsection{Number of $n$-equivalence classes}\label{subseq:nbrEquiv}
	A consequence of Propositions \ref{isometryImplyMonoEquiv} and  \ref{EqImpliesIso} is that, $n$-equivalent binomials give rise practically to the same trinomial codes. Hence the number of $n$-equivalence classes serves as an upper bound to the number of trinomials one should consider when searching for new trinomial codes.  To get the number of $n$-equivalence classes on $B$, by Corollary \ref{equivIsLocal}, it suffices to compute the number of classes on $B_k$ for each $0<k<n$. And from Remark \ref{nbrEqIsIndex}, we see that this number is closely related to the cardinal of $H_k=\{\alpha^{n-k}x^k + \alpha^n ~|~ \alpha \in R^\times\}.$ 
	On the other hand, from Equation (\ref{eqGeneralFormUnitGroup}) we see that $R^\times$  is a direct product of $T^{*}$ which is a cyclic group of order $p^{r}-1$ where $T$ is the Teichm\"uller set of $R$ and of $1+uR$ which is a $p$-group of order $p^{r(s-1)}$. Hence, according to the fundamental theorem of finite Abelian groups 
	we have that
	\begin{align}\label{URStructure}
		R^{\times} \simeq \left[\bigoplus^J_{j = 1} \mathbb{Z}_{p^{m_{j}}}\right]\oplus \mathbb{Z}_{p^{r}-1}
	\end{align}
	
	\noindent for some integer $J \geq 1$, such that for all $1\leq j\leq J$, $m_{j}\in \mathbb{N}$, $m_1\leq m_2\leq \ldots\leq m_J$  and $\prod\limits^J_{j=1}p^{m_{j}}=p^{r(s-1)}$, i.e. $\sum\limits^J_{j=1}m_j=r(s-1)$.
	The following two results will be helpful.
	\begin{lemma}\label{divisorOrderCyclic}
		Let $G$ be a cyclic group of order $r$ and let $\xi$ be a generator of $G$. 
		\begin{itemize}
			\item[(i)] For each divisor $q$ of $r$, $G$ has a unique subgroup of order $q$. 
			\item[(ii)] The elements of $G$ of order $q$ are the elements  $\displaystyle \xi^{\frac{sr}{q}}$ for each integer $s<q$ such that $\gcd(s,q)=1,$ and the number of these elements is $\Phi(q)$, where $\Phi$ denotes Euler's function.
		\end{itemize}
		
	\end{lemma}
	
	\begin{lemma}\label{orderProduct}
		Let $G_1,G_2,\ldots,G_n$ be finite abelian groups. The order of an element $(g_1,g_2,\ldots,g_n) \in G_1\times G_2\times\ldots\times G_n$ is the least common multiple of the orders of each $g_i$ in $G_i$, $1\leq i\leq n.$
	\end{lemma}

	Let us compute the cardinal of $H_k$ for an integer $0<k<n.$ Consider the map \begin{align*}
		\begin{array}{cccc}
			\theta:& R^\times & \longrightarrow & H_k  \\&&&\\ 
			& \alpha & \longrightarrow & \alpha^{n-k}x^k + \alpha^n
		\end{array} 
	\end{align*}
	
	\noindent$\theta$ is obviously a surjective group homomorphism of kernel $\mathrm{ker}(\theta)=\{\alpha \in R^\times ~|~\alpha^n=\alpha^{n-k}=1\}.$ Hence for $\alpha \in R^\times,$ $\alpha$ is in $\mathrm{ker}(\theta)$ if and only if $\alpha^n=\alpha^{n-k}=1,$ if and only if $\alpha^n=\alpha^{k}=1,$ if and only if the order of $\alpha$ in $R^\times$ divides $\mathrm{gcd}(n,k).$
	Let $D$ denote the set of positive divisors of $\mathrm{gcd}(n,k).$ The cardinal of $\mathrm{ker}(\theta)$ is the number of elements of $R^\times$ for which the order in $R^\times$ is an element of $D.$ 
	According to Equation (\ref{URStructure}), an element $\alpha \in R^\times$ can be represented as a $(J+1)$-tuple $(\alpha_1,\ldots,\alpha_J,\alpha_{J+1})$ where $\alpha_j \in \mathbb{Z}_{p^{m_{j}}}$ for all $1\leq j\leq J$ and $\alpha_{J+1} \in \mathbb{Z}_{p^{r}-1}.$ For $\alpha \simeq (\alpha_1,\ldots,\alpha_J,\alpha_{J+1})$ we will say that the order of $\alpha$ is of type $(\mathrm{ord}(\alpha_1),\ldots,\mathrm{ord}(\alpha_J),\mathrm{ord}(\alpha_{J+1}))$, where $\mathrm{ord}(.)$ denotes the order of the element.  By Lemma \ref{orderProduct}, the order of $\alpha$ is $\mathrm{lcm}(\mathrm{ord}(\alpha_1),\ldots,\mathrm{ord}(\alpha_J),\mathrm{ord}(\alpha_{J+1}))$, but as $\mathrm{gcd}(p,p^r-1)=1$ and $\alpha_{J+1} \in \mathbb{Z}_{p^{r}-1},$ we have $$\mathrm{ord}(\alpha)=\mathrm{lcm}(\mathrm{ord}(\alpha_1),\ldots,\mathrm{ord}(\alpha_J))\mathrm{ord}(\alpha_{J+1})=\mathrm{max}\{\mathrm{ord}(\alpha_1),\ldots,\mathrm{ord}(\alpha_J)\}\mathrm{ord}(\alpha_{J+1}).$$ Hence the order of an element $\alpha \in R^\times$ is of the form $p^lu,$ where $u$ is a positive divisor of $p^{r}-1$ and $l$ is an integer less or equal than $\mathrm{max}\{m_1,\ldots,m_J\}=m_J.$
	
	\noindent Let $d=p^lu$  where $u$ is a positive divisor of $p^{r}-1$ and $l$ is an integer less or equal than $m_J.$ For $\alpha \simeq (\alpha_1,\ldots,\alpha_J,\alpha_{J+1})$ to be of order $d,$ we must have $\mathrm{ord}(\alpha_{J+1})=u$ and $\mathrm{max}\{\mathrm{ord}(\alpha_1),\ldots,\mathrm{ord}(\alpha_J)\}=p^l$ which amounts to the following, $\mathrm{ord}(\alpha_{J+1})=u$ and for all $1\leq j\leq J,$ $\mathrm{ord}(\alpha_j) \leq \mathrm{min}\{p^l,p^{m_j}\}$ with $\mathrm{ord}(\alpha_i) = p^l$ for some integer $1\leq i\leq J.$  For $1\leq j\leq J$ let $l_j:=\mathrm{min}\{l,m_j\},$ and let $\varepsilon(l)$ be the smallest integer between $1$ and $J$ such that $l_{\varepsilon(l)}=l.$ Note that such $\varepsilon(l)$ exists because $l\leq m_J,$ and as $m_1\leq m_2\leq \ldots\leq m_J,$ we have $l_1\leq \ldots\leq l_{\varepsilon(l)-1}< l_{\varepsilon(l)}=l_{\varepsilon(l)+1}=\ldots=l_J=l.$
	
	\noindent The number of elements of $R^\times$ of order $d=p^lu$, is the sum of all the numbers of elements whose order is of type $(p^{t_1},p^{t_2},\ldots,p^{t_J},u)$ where for $1\leq j\leq J,$ $0\leq t_j \leq l_j$ and for some $\varepsilon(l)\leq i \leq J,$ $t_i=l.$ Using Lemma \ref{divisorOrderCyclic} \textit{(ii)}, the number of elements whose order is of type $(p^{t_1},p^{t_2},\ldots,p^{t_J},u)$ is $\displaystyle \Phi(u)\prod_{j=1}^{J}\Phi(p^{t_j}).$

	\noindent Hence, the number of elements of $R^\times$ of order $d=p^lu$, which we will denote $\#\mathrm{ord}(l,u)$, is 
	\begin{equation}
		\#\mathrm{ord}(l,u)=\mathlarger{\mathlarger{\sum}}_{\substack{0\leq t_i \leq l_i \\  1\leq i \leq J \\ t_{\omega}=l \text{ for some }\omega\geq \varepsilon(l)}}\Phi(u)\prod_{j=1}^{J}\Phi(p^{t_j})
	\end{equation}
	To simplify the previous equation, we consider the two sets 
	\begin{equation*}
		\begin{cases}
			T=\left\{(t_1,t_2,\ldots,t_J) ~|~ \forall~ 1\leq i \leq J, ~ 0\leq t_i \leq l_i\right\}, \\
			T'=\left\{(t_1,t_2,\ldots,t_J) ~|~ 
			0\leq t_i \leq l_i, \text{ for } 0\leq i\leq \varepsilon(l)-1, \text{ and }
			0\leq t_i \leq l-1, \text{ for } \varepsilon(l) \leq i \leq J
			\right\}.
		\end{cases}
	\end{equation*}
	We have then,  for $l>0$
	\begin{equation}\label{nbrEltOrd}
		\#\mathrm{ord}(l,u)=\mathlarger{\mathlarger{\sum}}_{(t_1,\ldots,t_J)\in T}\Phi(u)\prod_{j=1}^{J}\Phi(p^{t_j}) ~- \mathlarger{\mathlarger{\sum}}_{(t_1,\ldots,t_J)\in T'}\Phi(u)\prod_{j=1}^{J}\Phi(p^{t_j})
	\end{equation}
	and $\#\mathrm{ord}(0,u)=\Phi(u).$
	
	Thus, the cardinal of $\mathrm{ker}(\theta)$ is the sum of $\#\mathrm{ord}(l,u)$ for all integers $l,u$ such that $u$ divides $ p^r-1,$ $l\leq m_J$ and $d=p^lu \in D.$ Hence, by denoting $\displaystyle \omega(k)=\mathrm{min}\{v_p(\mathrm{gcd}(n,k)),m_J\}$ where $v_p()$ is the $p$-adic valuation, we get 
	\begin{align}\label{cardinalKerPhi}
		|\mathrm{ker}(\theta)| = \mathlarger{\sum}_{\substack{0\leq l \leq \omega(k) \\ u \mid \mathrm{gcd}(n,k,p^r-1)}} \#\mathrm{ord}(l,u)
	\end{align}
	Now that we have computed the cardinal of $\mathrm{ker}(\theta),$ the cardinal of $H_k$ follows easily. Indeed, using the first theorem of isomorphism we get 
	\begin{equation}\label{cardinalHk}
		|H_k|=\frac{|R^\times|}{|\mathrm{ker}(\theta)|}=\frac{p^{r(s-1)}(p^r-1)}{|\mathrm{ker}(\theta)|}
	\end{equation}
	The following theorem recapitulates the above discussion and provides the number of $n$-equivalence classes of trinomial codes over $R$.
	\begin{theorem}\label{nbrEquivClass}
		For an integer $0<k<n$, the number of of $n$-equivalence classes on $B_k$ is 
		\begin{equation}
			|B_k/\sim_n|=p^{r(s-1)}(p^r-1)\mathlarger{\mathlarger{\sum}}_{\substack{0\leq l \leq \omega(k) \\ u \mid \mathrm{gcd}(n,k,p^r-1)}} \#\mathrm{ord}(l,u)
		\end{equation}
		where $\displaystyle \omega(k)=\mathrm{min}\{v_p(\mathrm{gcd}(n,k)),m_J\}$ and $\#\mathrm{ord}(l,u)$ is as defined in Equation \ref{nbrEltOrd}.
	\end{theorem} 
	\begin{proof}
		For each integer  $0<k<n$, from Remark \ref{nbrEqIsIndex}, we see that the number of classes of equivalence of $\sim_n$ over $B_k$ is the index of $H_k,$ i.e. \begin{equation}
			|B_k/\sim_n|=[B_k : H_k]=\frac{|B_k|}{|H_k|}
		\end{equation}
		
		Moreover, we have that the cardinal of $B_k$ equals that of $(R^\times)^2,$ i.e. $p^{2r(s-1)}(p^r-1)^2.$ And, form equations (\ref{cardinalHk}) and (\ref{cardinalKerPhi}) we get that 
		\begin{equation}
			|H_k|=	{p^{r(s-1)}(p^r-1)}\bigg/
			\left({\sum_{\substack{0\leq l \leq \omega(k) \\ u \mid \mathrm{gcd}(n,k,p^r-1)}} \#\mathrm{ord}(l,u)}\right)
		\end{equation}
		where $\#\mathrm{ord}(l,u)$ is as defined in equation (\ref{nbrEltOrd}), and the result follows.
		
	\end{proof}
	\begin{remark}
		For $0<k<n,$ as $\gcd(n,k)=\gcd(n,n-k),$ the number of $n$-equivalence classes on $B_k$ equals that on $B_{n-k}.$ Moreover, the number of $n$-equivalence classes on $B$ follows immediately from Theorem \ref{nbrEquivClass} and Corollary \ref{equivIsLocal}.
	\end{remark}

	\begin{corollary}
		Let $k$ be  an integer such that $0<k<n.$ 
		\begin{itemize}
			\item[(i)] If $\mathrm{gcd}(n,k)=1,$ then there are $p^{r(s-1)}(p^r-1)$ $n$-equivalence classes on $B_k.$
			\item[(ii)] If $\mathrm{gcd}(n,p)=1$ or $\mathrm{gcd}(k,p)=1,$ then there are $p^{r(s-1)}(p^r-1)\mathrm{gcd}(n,k,p^r-1)$ $n$-equivalence classes on $B_k.$
			\item[(iii)] In particular when $\mathrm{gcd}(n,p)=1$ the number of $n$-equivalence classes on $B$ is $$ p^{r(s-1)}(p^r-1)\sum_{k=1}^{n-1}\mathrm{gcd}(n,k,p^r-1).$$ 
		\end{itemize}
	\end{corollary}
	\begin{proof}$\quad $
		\begin{description}
			\item[$(i)$] As $\gcd(n,k)=1,$ $\gcd(n,k,p^r-1)=1$ and $\omega(k)=\mathrm{min}\{v_p(\mathrm{gcd}(n,k)),m_J\}=0.$ Hence the result follows from Theorem \ref{nbrEquivClass} and the fact that $\#\mathrm{ord}(0,1)=1.$ 
			\item[$(ii)$] If $\mathrm{gcd}(n,p)=1$ or $\mathrm{gcd}(k,p)=1,$ then $v_p(\gcd(n,k))=0$ which implies that  $\omega(k)=0.$ Hence from Theorem \ref{nbrEquivClass} and the fact that $\#\mathrm{ord}(0,u)=\Phi(u)$, the number of $n$-equivalence classes on $B_k$ is $p^{r(s-1)}(p^r-1)\sum_{\substack{u \mid \mathrm{gcd}(n,k,p^r-1)}} \Phi(u)=p^{r(s-1)}(p^r-1)\mathrm{gcd}(n,k,p^r-1).$
			\item[$(iii)$] Follows immediately from $(ii)$ and Corollary \ref{equivIsLocal}.
		\end{description}
	\end{proof}
	
	\begin{remark}
		In the case of $R=\mathbb{F}_q$, $q$ being a power of a prime $p$, we have, for  each integer $0<k<n$, $\omega(k)=0$, hence Theorem \ref{nbrEquivClass} gives us that the number of $n$-equivalence classes on $B_k$ is 
		\begin{align}
			|B_k/\sim_n|=(q-1)\underset{ u \mid \mathrm{gcd}(n,k,q-1)}\sum \#\mathrm{ord}(0,u) \nonumber\\
			= (q-1)\underset{ u \mid \mathrm{gcd}(n,k,q-1)}\sum \Phi(u) \nonumber \\
			= (q-1)\mathrm{gcd}(n,k,q-1)
		\end{align}
		and the number of $n$-equivalence classes on $\mathbb{F}_q$ as a whole is $\displaystyle (q-1)\sum_{k=1}^{n-1}\mathrm{gcd}(n,k,q-1).$
	\end{remark}
	
	\begin{example}\textbf{(Trinomial codes over a Galois ring)}\\
		Let $R=GR(p^m,r)$ the Galois ring of characteristic $p^m$ and rank $r.$ The cardinal of the group of units of $R$ is $|R^\times|=p^{r(m-1)}(p^r-1).$ The following characterization of $R^\times$ is provided in  \cite[Theorem XVI.9]{mcdonald1974finite}.
		\begin{align}
			R^\times \simeq
			\begin{cases}
				\displaystyle	\mathbb{Z}_{p^r-1} \oplus \mathbb{Z}^r_{p^{m-1}}, \text{ ~~~~  if $p$ is odd, or if $p = 2$ and $m\leq 2$.}\\
				\displaystyle\mathbb{Z}_{2^r-1} \oplus \mathbb{Z}_2\oplus \mathbb{Z}_{2^{m-2}}\oplus \mathbb{Z}^{r-1}_{2^{m-1}}, \text{~~~~~~if $p = 2$ and $m> 2$.}
			\end{cases}
		\end{align}
		
		Let us consider for example trinomial codes of length $p^n-1$ over $GR(p^m,r)$ for an odd prime $p$ and integers $n,m,r \in \mathbb{N}^*$ such that $\gcd(n,r)=1.$ For $0<k<p^n-1,$ the number of $(p^n-1)$-equivalence classes on $B_k$ is $p^{r(m-1)}(p^r-1)\gcd(p^n-1,k,p^r-1)=p^{r(m-1)}(p^r-1)\gcd(p-1,k),$ which is a significant reduction from $(p^{r(m-1)}(p^r-1))^2$ the number of elements of $B_k.$	
	\end{example}
	\begin{example}\textbf{(Number of $n$-equivalence classes over the ring $\mathbb{F}_q[u]/\langle u^s\rangle$)}\\
		The ring $R=\mathbb{F}_q[u]/\langle u^s\rangle$ where $q=p^r,$ has invariants $(p,1,r,s,s).$ Using \cite[Proposition 2.7]{HOU200320}, we have 
		\begin{equation}
			R^\times \simeq\mathbb{Z}_{p^r-1} \oplus \left[\bigoplus_{i\in I} \left(\mathbb{Z}_{p^{\alpha(i)}}\right)^{r}\right]
		\end{equation}
		where $I=\displaystyle \left\{i\in\mathbb{Z}~ : ~1\leq i< s \text{ and } p\nmid i\right\},$ and for all $i\in I,$ $\displaystyle \alpha(i)=\lceil\log_{p}(\frac{s}{i})\rceil.$ \\
		Take for example $R=\mathbb{F}_{3^2}[u]/\langle u^4\rangle.$  Its group of units is isomorphic to $\mathbb{Z}_{8} \oplus \mathbb{Z}_3^2\oplus \mathbb{Z}_{3^2}^2.$ Consider trinomial codes of length $12$ over $R.$ We have three cases:
		\begin{itemize}
			\item[(i)] For each $0<k<12$ such that $k \not\equiv 0 \pmod{3}$: the number of $12$-equivalence classes on $B_k$ is $5832\times \gcd(k,4).$
			\item[(ii)] As $\omega(k)=\mathrm{min}\{v_3(\gcd(12,k)),2\},$ we have $\omega(3)=\omega(6)=\omega(9)=1,$ moreover, as $\gcd(12,3,8)=\gcd(12,9,8)=1,$ the number of $12$-equivalence classes on $B_3$ is the same as on $B_9$ and after calculation we get $472392.$
			\item[(iii)]  For $k=6,$ we have $\gcd(12,6,8)=2,$ and as $\Phi(1)=\Phi(2)=1,$ we have $|B_6/\sim_{12}|=2\times |B_3/\sim_{12}|.$ 
		\end{itemize}
		Hence, the number of $12$-equivalence classes on $R$ as a whole, which is the sum of the numbers of classes on each $B_k$ for $0<k<12,$ is $1982880.$ Notice that it is significantly smaller than the number of elements of $B$ which equals $11\times |B_0|=11\times |R^\times|^2=374134464.$
		
	\end{example}
	
	\begin{example}\textbf{(Trinomial codes over a finite field)}\\
		For two primes $p$ and $q$ and positive integers $n$ and $m,$ we consider the  case of trinomial codes of length $p^n$ over $\mathbb{F}_{q^m}.$  For $0<k<p^n,$ the number of $p^n$-equivalence classes over $B_k$ is given by
		$ (q^m-1)\times \gcd(p^n,k,q^m-1).$ If $p=q$ or $k \not\equiv 0 \pmod{p}$ or $k \equiv 0 \pmod{q}$ then the number of $p^n$-equivalence classes over $B_k$ reduces to only $q^m-1$.\\
		Take for example $p=3,~q=2,~n=3\text{ and }m=2.$ For $0<k<27,$ we distinguish between two cases:
		
		\begin{enumerate}
			\item If $k \equiv 0 \pmod{3}$: 
			The number of $27$-equivalence classes over $B_k$ is $9$. In this case, we consider all pairs $(a, b)$ from $\mathbb{F}_4^* \times \mathbb{F}_4^*$.
			
			\item If $k \not\equiv 0 \pmod{3}$: 
			The number of $27$-equivalence over $B_k$ classes becomes $3$.			
		\end{enumerate}
		Hence over the field $ \mathbb{F}_4,$ instead of considering all 234 classes of trinomial codes of length $27$, we only need to study 126, the number of $27$-equivalence classes over $ \mathbb{F}_4.$
		
	\end{example}
	
	\section{Trinomials of the form $x^n-a_1x-a_0$}\label{sec:Tri1}
	In this section, we focus on polycyclic codes associated to trinomials of the form $x^n-a_1x-a_0$ where $a_1$ and $a_0$ are units in $R^\times$. 
	We will need the next lemma.
	
	\begin{lemma}(see \cite{mihet2010legendre})\label{mihet}
		
		For an integer $N>1$, the greatest common divisor of $\displaystyle \binom{N}{1},\binom{N}{2},\ldots.,\binom{N}{N-1},$ is either $1$, if $N$ is not a power of a prime, or a prime $p$ if $N$ is a power of $p$.
		
	\end{lemma}
	
	\begin{theorem}\label{charpmIsoIsEquiv}
		If the characteristic of the ring $R$ is $p^m$ with $m>1$, then for any two elements $a(x)$ and $b(x)$ of $B_1$, we have the following equivalence: $a(x)$ and $b(x)$ are $n$-isometric if and only if $a(x)$ and $b(x)$ are $n$-equivalent.  
	\end{theorem}
	\begin{proof}
		
		Let $a(x)=a_1x+a_0$ and $b(x)=b_1x+b_0$, in $B_1$. If $a(x) \cong_{n} b(x)$, let $\phi: R[x]/ \langle x^{n} - a(x) \rangle \rightarrow R[x]/ \langle x^{n} - b(x) \rangle$ be an associated isometry.
		By Proposition \ref{isoForm}, there exists an element $\alpha$ in $R^\times$ and an integer $0< j \leq n-1$ such that $\displaystyle \phi(x)=\alpha x^{j}$.\\
		We have
		\begin{equation*}
			\phi(a(x))=a_1\phi(x)+a_0=a_1\alpha x^j+a_0\mod (x^{n} - b(x))
		\end{equation*}
		and 
		\begin{align*}
			\phi(a(x))=\phi(x^n)=\phi(x)^n=\alpha^n x^{nj}=\alpha^n(b_1x+b_0)^j=\alpha^n\sum_{l=0}^{j}\binom{j}{l}b_1^lx^lb_0^{j-l}=\\ =\alpha^nb_1^jx^j + \alpha^n\sum_{l=1}^{j-1}\binom{j}{l}b_1^lx^lb_0^{j-l} + \alpha^nb_0^j \mod (x^{n} - b(x))
		\end{align*}
		
		\noindent As $\alpha, a_1$ and $b_1$ are units and $j<n$, by the uniqueness of the representative polynomial of degree strictly less than $n$ for each class modulo $(x^{n} - b(x)),$ we must have in $R[x]$
		\begin{equation*}
			\displaystyle
			a_1\alpha x^j+a_0=\alpha^nb_1^jx^j + \alpha^n\sum_{l=1}^{j-1}\binom{j}{l}b_1^lx^lb_0^{j-l} + \alpha^nb_0^j
		\end{equation*}
		hence $\displaystyle\binom{j}{1}=\binom{j}{2}=\ldots.=\binom{j}{j-1}=0$ in $R$, which means that the characteristic $p^m$ of $R$ divides $\displaystyle \binom{j}{i}$ for all $1\leq i\leq j-1,$ hence $p^m$ divides $\displaystyle \mathrm{gcd}\left(\binom{j}{1},\binom{j}{2},\ldots.,\binom{j}{j-1}\right),$ but from Lemma \ref{mihet}, as $m>1,$ that cannot be the case for $j>1$. Hence $j=1$ and 
		\begin{align*}
			\begin{array}{cccc}
				\phi:& {R[x]}/ \langle x^{n} -a(x) \rangle & \longrightarrow & {R[x]}/ \langle x^{n} - b(x) \rangle  \\&&&\\ 
				& f(x) & \longrightarrow & f(ax)
			\end{array} 
		\end{align*}
		is an $R$-algebra isomorphism, i.e. $a(x)\sim_n b(x)$.\\
		The converse implication follows from Proposition \ref{EqImpliesIso}.
		
	\end{proof}
	
	\begin{theorem}\label{charpIsoImpliesEquiv}
		If $R$ is of characteristic a prime $p,$ then for any two elements $a(x)$ and $b(x)$ in $B_1$, if $a(x)$ and $b(x)$ are $n$-isometric then either $a(x)$ and $b(x)$ are $n$-equivalent, or there exists an integer $l\in \mathbb{N}^*$ such that $a(x)$ and $b(x)^{\star p^l}$ are $n$-equivalent and $p^l<n$, where  $b(x)^{\star p^l}=\underbrace{b(x)\star b(x)\star \ldots\star b(x)}_\text{$p^l$ times}.$
	\end{theorem}
	\begin{proof}
		Let $a(x)=a_1x+a_0$ and $b(x)=b_1x+b_0$, in $B_1$. If $a(x) \cong_{n} b(x)$, let $\phi: R[x]/ \langle x^{n} - a(x) \rangle \rightarrow R[x]/ \langle x^{n} - b(x) \rangle$ be an associated isometry. From Proposition \ref{isoForm} and following the proof of Theorem \ref{charpmIsoIsEquiv}, there exists an element $\alpha$ in $R^\times$ and $0< j \leq n-1$ such that $\displaystyle \phi(x)=\alpha x^{j}$, and  $p$ divides $\displaystyle \mathrm{gcd}\left(\binom{j}{1},\binom{j}{2},\ldots.,\binom{j}{j-1}\right).$ Hence, by Lemma \ref{mihet}, $j$ must be either $1$ or a power of $p$. If $j=1$ then $a(x) \sim_n b(x).$ Suppose there exists an integer $l \in \mathbb{N}^*$ such that $j=p^l.$ We have 
		\begin{equation*}
			\phi(a(x))=a_1\phi(x)+a_0=a_1\alpha x^{p^l}+a_0\mod (x^{n} - b(x))
		\end{equation*}
		and 
		\begin{align*}
			\phi(a(x))=\phi(x^n)=\phi(x)^n=\alpha^n x^{np^l}=\alpha^n(b_1x+b_0)^{p^l}=\alpha^nb_1^{p^l}x^{p^l}+\alpha^nb_0^{p^l} \mod (x^{n} - b(x)) ~~~
		\end{align*}
		\noindent hence, as $\alpha, a_1$ and $b_1$ are units and $p^l=j<n$, by uniqueness of the representative polynomial of degree strictly less than $n$ for each class modulo $(x^{n} - b(x)),$ we have $a_1=\alpha^{n-1}b_1^{p^l}$ and $a_0=\alpha^nb_0^{p^l},$  which by Theorem \ref{equivCharac}, implies that $a(x)\sim_n b(x)^{\star p^l}.$
		
	\end{proof}
	
	\begin{corollary}
		For all $a(x)=a_1x+a_0\in B(x)$, 
		$a(x)$ is $n$-isometric to $x+1$ if and only if there exists $\alpha \in R^\times$ such that $a_1=\alpha^{n-1}$ and $a_0=\alpha^n$.
	\end{corollary}
	\begin{proof}
		From Theorem \ref{charpmIsoIsEquiv} and Theorem \ref{charpIsoImpliesEquiv}, we get that for every finite chain ring $R$, $a(x) \cong_n (x+1)$ if and only if $a(x) \sim_n (x+1),$ and the result follows from the characterization of Theorem \ref{equivCharac}.
	\end{proof}
	
	\section{Equivalence of some additive trinomial codes}\label{sec:additive}
	
	Additive codes are a straightforward generalization of linear codes that have become of great interest due to their useful applications in quantum error-correcting codes. They have been extensively studied recently using different approaches and over different alphabets, in particular, additive codes over finite rings have gathered a lot of attention in recent years \cite{Otal,samei2017cyclic, mahmoudi2019additive}. In this section, we generalize the $n$-equivalence relation defined in Section \ref{eq} to some cases of trinomial additive codes over a finite chain ring $R$.
	
	Let $S\subset R$ be a subring of $R$. A nonempty set $C$ of $R^n$ is called an \textit{$S$-linear} (or \textit{$S$-additive}) code of length $n,$ if it is an $S$-submodule of $R^n.$ A \emph{monomial transformation} $\tau : R^n \rightarrow R^n $ is an $S$-linear automorphism of the form $\tau(c_0,...,c_{n-1})=(\psi_0(c_{\sigma(0)}),...,\psi_{n-1}(c_{\sigma(n-1)}))$ where $\sigma$ is a permutation of $\{0,\ldots,n-1\}$ and $\psi_0,\ldots,\psi_{n-1} \in \mathrm{Aut}_S(R)$ are $S$-module automorphisms of $R.$ Note that, as $\mathrm{Aut}_R(R)$ is the units group $R^n$ acting on $R$ by multiplication, we end up with the same definition of a monomial transformation in the case of linear codes over $R.$		
	Two $S$-linear codes $C_1,C_2 \subset R^n$ are said to be \textit{monomially equivalent} if there exists a monomial transformation $\tau$ such that $C_1=\tau(C_2).$
	\begin{definition}
		Let $C$ be an $S$-linear code over $R$ and let $a=(a_0,a_1,...,a_{n-1})\in S^n.$ $C$ is called an additive polycyclic code induced by $a,$ if for any $ c=\left(c_{_0},c_{_1},\ldots, c_{_{n-1}}\right)\in C $ we have $ \left(0,c_{_0},\ldots, c_{_{n-2}}\right)+c_{_{n-1}}\left(a_{_0},a_{_1},\ldots, a_{_{n-1}}\right) \in C.$
	\end{definition}
	Using the polynomial representation of codes, the next lemma is immediate.
	\begin{lemma}\label{additivePolyChara}
		Let $S\subset R$ and $C$ an $S$-linear code over $R.$ Then $C$ is an additive polycyclic code of length $n$ and induced by the polynomial $a(x)\in S[x],$ if and only if $C$ is an $S[x]$-submodule of $R[x]/\langle x^n -a(x)\rangle.$
	\end{lemma}
	
	\subsection{Restriction to the Teichm\"uller set}
	
	Following the notations of Theorem \ref{fcrAsGrQuotient}, let  $\displaystyle R={\mathrm{GR}(p^{m},r)[u]}/{\langle g(u),p^{m-1}u^{t}\rangle}.$ We start by considering $\mathrm{GR}(p^{m},r)$-linear codes over $R.$ 
	To classify $\mathrm{GR}(p^{m},r)$-additive trinomial codes over $R,$ we can make use of the $n$-equivalence relation we defined earlier. Indeed, if we choose the coefficient $\alpha$ in Definition \ref{def:n} as an element of $\mathrm{GR}(p^{m},r)^\times,$ the $R$-algebra isomorphism defined is in fact a $\mathrm{GR}(p^{m},r)$-algebra isomorphism that relates trinomials with coefficients in $\mathrm{GR}(p^{m},r)^\times.$ In particular, as the nonzero elements of the  Teichm\"uller set $T$ of $R$ form a subgroup of $\mathrm{GR}(p^{m},r)^\times,$ we consider the following.	Let us define $B_T,$ $B_{Tk}$ and $H_{Tk}$ as $B_T=T^*[x]\cap B,$  $B_{T,k}=T^*[x]\cap B_k$ and $H_{T,k}=T^*[x]\cap H_k$ respectively.

	\begin{definition}\label{def:nT}
		Let $a(x),b(x)\in B_T$. We say that $a(x) \text{ and }b(x)$ are $(n,T)$-equivalent and denote $a(x) \sim_{n,T} b(x)$, if there exists an element $\alpha$ in $T^*$ such that 
		\begin{align*}
			\begin{array}{cccc}
				\phi:& {R[x]}/ \langle x^{n} -a(x) \rangle & \longrightarrow & {R[x]}/ \langle x^{n} - b(x) \rangle  \\&&&\\ 
				& f(x) & \longrightarrow & f(\alpha x)
			\end{array} 
		\end{align*}
		is an $R$-algebra isomorphism.
	\end{definition}
	
	\begin{proposition}
		Let $a(x),b(x)\in B_T$ such that $a(x)\sim_{n,T} b(x)$ and $\alpha\in T^*$ such that $\phi : x \rightarrow \alpha x$ is an associated isomorphism. Let $C$ be a $\mathrm{GR}(p^{m},r)$-linear $(x^n-a(x))$-polycyclic code of length $n$ over $R$. Then $\phi(C)$ is a $\mathrm{GR}(p^{m},r)$-linear $(x^n-b(x))$-polycyclic code of length $n$ over $R,$ and $C$ and $\phi(C)$ are monomially equivalent.
	\end{proposition}
	\begin{proof}
		Using Lemma \ref{additivePolyChara}, it suffices to prove that $\phi(C)$ is a $\mathrm{GR}(p^{m},r)[x]$-submodule of ${R[x]}/ \langle x^{n} - b(x) \rangle$ for it to be a $\mathrm{GR}(p^{m},r)[x]$-linear $(x^n-b(x))$-polycyclic code. $\phi(C)$ is obviously an additive subgroup of ${R[x]}/ \langle x^{n} - b(x) \rangle$.  For $g(x)\in \mathrm{GR}(p^{m},r)[x]$ and $c(x)\in C$ we have $g(x)\phi(c(x))=\phi(g(\alpha^{-1}x)c(x)),$ and as $\alpha \in T^*\subset\mathrm{GR}(p^{m},r),$ $g(\alpha^{-1}x)\in \mathrm{GR}(p^{m},r)[x],$ hence, as $C$ is a $\mathrm{GR}(p^{m},r)[x]$-module, we have $g(\alpha^{-1}x)c(x)\in C$ and thus $g(x)\phi(c(x))\in \phi(C).$ Moreover, $\phi$ is obviously a monomial transformation, indeed for $(c_0,c_1,...,c_{n-1})\in R^n,$ $\phi(c_0,c_1,...,c_{n-1})=(c_0,\alpha c_1,...,\alpha^{n-1}c_{n-1})$ and $1,\alpha,...,\alpha^{n-1} \in \mathrm{GR}(p^{m},r)^\times.$ So $C$ and $\phi(C)$ are monomially equivalent.
	\end{proof}

	It is immediate that for all $a(x),b(x)\in B_T$ we have that if  $a(x) \text{ and }b(x)$ are $(n,T)$-equivalent then they are $n$-equivalent. Moreover, from the fact that $T^*$ is a subgroup of $R^\times$, we have that $\sim_{n,T}$ is an equivalence relation on $B_T,$ $B_{Tk}$ is a subgroup of $(B_k,\star)$ and $H_{Tk}$ is a subgroup of $H_k$. Hence, numerous properties of $(n,T)$-equivalence can be immediately deduced from Section \ref{eq}. We regroup some of them in the next proposition for convenience. 
	\begin{proposition}
		Let $a(x),b(x) \in B_T.$
		\begin{enumerate}
			\item[(i)] If $a(x) \sim_{n,T} b(x)$ then there exists an integer $0<k<n$ such that $a(x),b(x) \in B_{Tk}.$
			\item[(ii)] For each $0<k<n$, the relation $``\sim_{n,T}"$ induces an equivalence relation on $B_k$ and we have
			$ \displaystyle
			B_T/\sim_n ~= \underset{{0<k<n}}{\mathlarger{\mathlarger{\sqcup}}}( B_{Tk}/\sim_n).
			$
			\item[(iii)] The number of classes of equivalence of $``\sim_{n,T}"$ over $B_T$ equals the sum of the numbers of classes over each $B_{Tk}$.
			\item[(iv)] For $0<k<n,$ if   $a(x)=a_1x^k+a_0$ and $b(x)=b_1x^k+b_0,$ then $a(x) \sim_{n,T} b(x)$ if and only if there exists an element $\alpha$ in $T^*$ such that $
			b_1\alpha^{n-k}-a_1=b_0\alpha^n-a_0=0.$ 
			\item[(v)] For $a(x)\in B_{Tk},$ its $(n,T)$-equivalence class is $a(x)\star H_{Tk}$ and its cardinal is that of $H_{Tk}.$
			\item[(vi)] The number of $(n,T)$-equivalence classes over $B_{Tk}$ is the index $[B_{Tk}:H_{Tk}]$ of the subgroup $H_{Tk}$ in $B_{Tk}.$
		\end{enumerate}
	\end{proposition}
	\begin{remark}\label{subgroupsOfRtimes}
		Note that the previous proposition holds true if we consider any other subgroup of $R^\times$ instead of $T^*.$
	\end{remark}
	As in the case of trinomial codes in Section \ref{eq}, the number of $(n,T)$-equivalence classes serves as an upper bound to the number of classes of $\mathrm{GR}(p^{m},r)$-linear trinomial codes over $R$ with different parameters. From the precedent proposition, we see that this number equals the index of $H_{Tk}$ in $B_{Tk}.$ As $T^*$ is cyclic of order $p^r-1,$ let  $T^*=\langle \xi\rangle$ for some element $\xi\in T^*.$ We have $\displaystyle H_{Tk}=T^*[x]\cap H_k=\{\alpha^{n-k}x^k + \alpha^n ~|~ \alpha \in T^*\}=\{\xi^{i(n-k)}x^k + \xi^{in} ~|~ 1\leq i\leq p^r-1\}=\langle \xi^{n-k}x^k + \xi^{n} \rangle.$ Hence the cardinal of $H_{Tk}$ is the order of  $\xi^{n-k}x^k + \xi^{n}$ in $(B_{Tk},\star)$ which is the same as the order of $(\xi^{n-k}, \xi^{n})$ in $\displaystyle (R^\times)^2.$ Thus we have,
	\begin{align}
		|H_{Tk}|=\mathrm{lcm}\left(\frac{p^r-1}{\gcd(p^r-1,n-k)},\frac{p^r-1}{\gcd(p^r-1,n)}\right)
		=\frac{p^r-1}{\gcd(p^r-1,k,n)}
	\end{align} 
	Finally, the number of $(n,T)$-equivalence classes on $B_{Tk}$ is 
	$\displaystyle {|B_{Tk}|}/{|H_{Tk}|}=(p^r-1)^2/{|H_{Tk}|}=(p^r-1)\gcd(p^r-1,k,n).$
	
	\begin{remark}
		Note that the number of $(n,T)$-equivalence classes on $B_{Tk}$ characterizes the codes with associated trinomials with coefficients in $T^*.$ If we consider the trinomial coefficients as elements in $\mathrm{GR}(p^{m},r)^\times$ instead, we will end up with $\displaystyle \frac{|\mathrm{GR}(p^{m},r)^\times|^2}{p^r-1}\cdot\gcd(p^r-1,k,n)=p^{2r(m-1)}(p^r-1)\cdot\gcd(p^r-1,k,n)$ classes.
	\end{remark}
	
	\subsection{Restriction to other subgroups of $R^\times$}
	
	We will consider a specific case of additive codes over a finite chain ring called \textit{Galois additive}. See \cite{Otal} for a detailed discussion about the structure of these codes and their characteristics. Following the notations of Theorem \ref{fcrAsGrQuotient}, let  $\displaystyle R={\mathrm{GR}(p^{m},r)[u]}/{\langle g(u),p^{m-1}u^{t}\rangle}$ and let $\omega \in R$ such that the Teichm\"uller set of $R$ is $T=\{0,1,\omega,\omega^2,...,\omega^{p^r-2}\}.$ Denote $S={\mathbb{Z}_{p^m}[u]}/{\langle g(u),p^{m-1}u^{t}\rangle}.$ We have  $R=S[\omega]$ and each element of $R$ has a unique representation as  $a_0 + a_1 \omega + \dots + a_{r-1}\omega^{r-1}$ where $a_i\in S.$ The precedent expression shows that $S^\times$  the group of units of $S$ is a subgroup of $R^\times.$ Hence, as pointed in Remark \ref{subgroupsOfRtimes}, we can define an equivalence relation \emph{$(n,S)$-equivalence} as in Definition \ref{def:nT} by replacing $T^*$ with $S^\times$ and the same results will follow.		
	Moreover, $S^\times$ is isomorphic to a direct sum of a cyclic group of order $p-1$ (which is the unique subgroup of $T^*$ of order $p-1$) and a $p$-group of order $p^{(m-1)e + t -1}$ (which is a subgroup of $1+uR$). Hence, using the same computations as in Subsection \ref{subseq:nbrEquiv}, we get the number of $(n,S)$-equivalence classes based on the decomposition of $S^\times$ into cyclic groups.
	Note that the same can be done for each subring of $R$ of the form ${\mathrm{GR}(p^{m},r')[u]}/{\langle g(u),p^{m-1}u^{t}\rangle}$ with $r'$ a positive divisor of $r,$ and in fact, $S$ is such a subring with $r'=1.$
	
	Proposition \ref{isometryImplyMonoEquiv}, which relates the notions of $n$-isometry and monomial equivalence, is a consequence of the fact that linear codes over finite chain rings have the MacWilliams extension property. Recall that the MacWilliams extension theorem states that over finite fields, two linear codes are monomially equivalent if and only if there exists a linear isomorphism between them that preserves the Hamming weight. In \cite{wood1999duality}, J. A. Wood  generalized this theorem to the case of finite Frobenius rings which finite chain rings are a subclass of. And in \cite{wood2009foundations}, numerous results concerning the extension problem for linear codes defined over
	finite modules are presented. Especially,  \cite[Theorem 6.2]{wood2009foundations} states that for a module $A$, if $A$ has the extension property with respect to the Hamming weight, then the socle of $A$ must be cyclic. 
	As $S$ is a finite ring, it is a known result (see \cite[15.17. Proposition]{anderson1992rings} for example) that the socle of an $S$-module $M$ is the annihilator of the Jacobson radical $J(S)$ of $S$ in $M.$ Thus, as an $S$-module, we have $\mathrm{soc}(R)=\{x\in R ~|~ yx=0 \text{ for all } y\in J(S)\}.$ As $R$ is a free $S$-module of basis $\{1,\omega, \omega^2,\ldots,\omega^{r-1}\},$ and as $S$ is a local principal ring with maximal ideal generated by an element $\gamma \in S$ of nilpotency index $s=(m-1)e-t,$  we have 
	\begin{align*}
		\mathrm{soc}(R)=&\left\{ \sum_{i=0}^{r-1}a_i\omega^i ~|~ \text{ for all } 0\leq i\leq r-1,~ a_i\in S \text{ and } a_i\gamma S =0\right\} \\=&\left\{\sum_{i=0}^{r-1}a_i\omega^i ~|~ \text{ for all } 0\leq i\leq r-1,~ a_i\in \gamma^{s-1}S\right\}\\
		=&\bigoplus_{i=0}^{r-1}\gamma^{s-1}S\omega^{i}.
	\end{align*}
	Hence the socle of $R$ as an $S$-module is not cyclic, which implies that $S$-linear codes over $R$ do not have the extension property w.r.t. the Hamming weight. A consequence of this is that unlike Proposition \ref{isometryImplyMonoEquiv}, an $R$-algebra isomorphism that preserves the Hamming weight does not necessarily induce a monomial equivalence between $S$-linear codes, which means that the $n$-isometry relation loses a powerful property when generalized to this case of additive codes.	Note however, that the $n$-equivalence relation induces a monomial transformation between additive codes as we have seen previously, and this is because the $R$-algebra isomorphism defined in the case of $n$-equivalence, is in itself a monomial transformation.\\

	\section*{Conclusion}
	
	In this paper, we investigated the equivalence between classes of polycyclic codes associated with trinomials of degree \( n \) over a finite chain ring \( R \). We introduced two equivalence relations, referred to as \( n \)-equivalence and \( n \)-isometry, which extend the notions of \( n \)-equivalence and \( n \)-isometry previously established for constacyclic codes. We derived a formula for the number of \( n \)-equivalence classes and provided conditions under which two families of trinomial codes are equivalent. Furthermore, we examined the relationship between \( n \)-equivalence and \( n \)-isometry in the specific case of trinomials of the form \( x^n - a_1x - a_0 \in R[x] \). Finally, we extended our results to certain trinomial additive codes over \( R \). As a direction for future research, we aim to explore these equivalence relations in the more general context of linear and additive polycyclic codes over finite chain rings.

	\printbibliography
\end{document}